\documentclass[twocolumn,%
amsmath,amssymb,aps,pra]{revtex4-2}

\usepackage{graphicx}
\usepackage{physics}
\usepackage{comment}
\usepackage{color}
\usepackage{dcolumn}
\usepackage{bm}

\usepackage[colorlinks,urlcolor=blue,citecolor=blue,linkcolor=blue]{hyperref}

\newcommand{\up}{\uparrow}
\newcommand{\down}{\downarrow}

\renewcommand{\k}{{\bf k}}
\newcommand{\p}{{\bf p}}

\newcommand{\q}{{\bf q}}

\newcommand{\0}{{\bf 0}}
\newcommand{\eq}{\epsilon_{\q}}
\newcommand{\ek}{\epsilon_{\k}}
\newcommand{\ep}{\epsilon_{\p}}

\newcommand{\ef}{E_{\rm F}}
\newcommand{\kf}{k_{\rm F}}
\newcommand{\eb}{\varepsilon_{\rm B}}
\newcommand{\as}{a_{\rm{s}}}

\newcommand{\nn}{\nonumber}

\newcommand{\replyadd}[1]{{\color{black}{#1}}}

\newcommand{\fourvec}[4]{\left[\begin{array}{cc} #1 & #2 \\ #3 & #4 \end{array} \right]}

\begin{document}

\title{Rabi oscillations and magnetization of a mobile spin-1/2 impurity in a Fermi sea}

\author{Brendan C. Mulkerin}
\affiliation{School of Physics and Astronomy, Monash University, Victoria 3800, Australia}
\affiliation{ARC Centre of Excellence in Future Low-Energy Electronics Technologies, Monash University, Victoria 3800, Australia}

\author{Jesper Levinsen}
\affiliation{School of Physics and Astronomy, Monash University, Victoria 3800, Australia}
\affiliation{ARC Centre of Excellence in Future Low-Energy Electronics Technologies, Monash University, Victoria 3800, Australia}

\author{Meera M. Parish}
\affiliation{School of Physics and Astronomy, Monash University, Victoria 3800, Australia}
\affiliation{ARC Centre of Excellence in Future Low-Energy Electronics Technologies, Monash University, Victoria 3800, Australia}

\date{\today}
\begin{abstract}
We investigate the behavior of a mobile spin-1/2 impurity atom immersed in a Fermi gas, where the interacting spin-$\up$ and non-interacting spin-$\down$ states of the impurity are Rabi coupled via an external field. This scenario resembles the classic problem of a two-state system interacting with a dissipative environment, but with an added dimension provided by the impurity momentum degree of freedom. In this case, the impurity can become ``dressed'' by excitations of the Fermi sea to form a Fermi polaron quasiparticle. For the steady-state system, where the impurity has thermalized with the medium, we derive exact thermodynamic relations that connect the impurity magnetization with quasiparticle properties such as the number of fermions in the dressing cloud. We show how the thermodynamic properties %
evolve with increasing Rabi coupling and we present exact analytical results in the limits of weak and strong Rabi coupling. 
For the dynamics of the Rabi-driven Fermi polaron, we formulate a theoretical approach based on correlation functions that respects conservation laws and allows the efficient calculation of Rabi oscillations for a range of time scales and impurity momenta beyond what has been achieved previously. Our results are in good agreement with recent experiments on the Rabi oscillations of the attractive polaron, and they reveal how the Rabi oscillations are influenced by the interplay between the polaron and its dressing cloud.

\end{abstract}

\maketitle
\section{Introduction}

The problem of a quantum impurity interacting with a Fermi medium has attracted much attention recently, owing to its importance in a wide variety of systems ranging from cold atomic gases~\cite{Schiro2009,nasc2009,Kohstall2012,Koschorreck2012,Zhang2012,Wenz2013,Cetina2015,Ong2015,Cetina2016,Scazza2017,Oppong2019,Yan2019,Ness2020,Fritsche2021} to doped semiconductors~\cite{Sidler2017,Efimkin2017}.
Most notably, it connects to the deeper questions of how a quantum system is affected by its environment, and how interacting many-body systems can be described using \emph{quasiparticles}---particles that resemble the bare non-interacting particles but with modified properties such as their charge or mass. 

For the case of a mobile quantum impurity, 
the bare impurity can become dressed by excitations 
of the background Fermi gas to form a so-called Fermi polaron quasiparticle~\cite{Massignan2014,Scazza2022,parish2023fermi}. 
Such Fermi polarons have been probed extensively in the context of cold atoms, where the coupling to an auxiliary internal spin state of the impurity atom has been used to access the energy spectrum~\cite{Schiro2009,Kohstall2012} as well as the coherent dynamics of quasiparticle formation and decay~\cite{Cetina2016}.  
A particularly straightforward and powerful experimental protocol is to continuously couple the non-interacting auxiliary impurity spin state with an impurity spin state that strongly interacts with the Fermi gas, thus driving Rabi oscillations that are sensitive to the polaron quasiparticle properties~\cite{Kohstall2012,Scazza2017,Oppong2019,Cetina2016}. This has allowed the precise determination of the quasiparticle residue (the squared overlap with the non-interacting state) and quasiparticle lifetime~\cite{Kohstall2012,Scazza2017,Oppong2019,Adlong2020}. 
However, an interesting and largely unexplored direction is the physics of the Rabi-coupled system itself, including how the Fermi polaron is changed by the Rabi coupling, %
as well as the magnetization in the steady state. \replyadd{This has only very recently been achieved in a homogeneous Fermi-gas experiment~\cite{vivanco2023}. The Rabi-coupled Fermi polaron} is also related to the problem of a spin-1/2 system in a dissipative environment~\cite{Leggett87,Knap2013}, but with the added dimension of an impurity momentum degree of freedom, \replyadd{thus allowing the existence of quasiparticles with non-zero overlap with the non-interacting state}.

In this paper, we theoretically investigate both the dynamics and the thermodynamics of the Rabi-coupled spin-1/2 impurity in a Fermi gas. 
We present exact thermodynamic relations that capture the interplay between the impurity magnetization and the impurity-medium interactions in the thermalized steady-state system. In the limit of weak Rabi coupling, we derive exact analytical results that connect the magnetization to the polaron energy and residue, and these are found to be remarkably robust with respect to increasing Rabi coupling. %
On the other hand, once the Rabi coupling far exceeds the Fermi energy of the medium, such that the impurity spin flips are faster than the medium's ability to respond,  we see the emergence of qualitatively different polaron quasiparticles with properties that can also be obtained analytically. 

For the quantum dynamics across a range of time scales, the full description of the Rabi oscillations is a challenging many-body problem, since it goes beyond standard theories based on linear response. 
To tackle this, we formulate a new theoretical approach based on correlation functions that respects conservation laws  and can describe the dynamics of the impurity and its polaronic dressing cloud.  %
Our approach is equivalent to previous variational methods~\cite{Parish2016,Adlong2020}, but is numerically much more efficient, allowing us to investigate thermalization as the system approaches the steady state, as well as the finite-temperature lifetime of the attractive polaron observed in experiment~\cite{Scazza2017}.

The paper is organized as follows. In Sec.~\ref{sec:model} we introduce our model of the spin-1/2 impurity in a Fermi gas, and discuss the impurity Green's function and spectral function that form the basis of our theoretical approach. We introduce the thermodynamic properties characterizing interactions and magnetization in the Rabi coupled system in Sec.~\ref{sec:thermodyn}, and we obtain their perturbatively exact expressions in the limit of weak and strong Rabi drives. Section~\ref{sec:approx} contains our numerical results for the thermodynamic properties at intermediate Rabi drive, as well as the impurity spectral function. Here we use a many-body $T$-matrix approximation, which becomes perturbatively exact in the limits of weak and strong Rabi drive. Finally, Sec.~\ref{sec:dynamics} contains our calculations of the impurity dynamics, which we obtain directly from the Green's function by introducing a two-body correlation function. In Sec.~\ref{sec:conc} we conclude. Additional technical details are in the appendices.

\section{Model and formalism}
\label{sec:model}

We consider an impurity atom with two internal hyperfine states $\sigma=\up,\down$ that is immersed in a Fermi gas. The total Hamiltonian consists of three terms, $\hat{H} = \hat{H}_0+\hat{H}_{\Omega}+\hat{V}$, where 
\begin{subequations}\label{eq:hamil}
\begin{alignat}{1}
\hat{H}_0 &= \sum_{\k}\left[\epsilon^{\phantom{\dagger}}_{\k}\hat{c}^{\dagger}_{\k\uparrow}\hat{c}^{\phantom{\dagger}}_{\k\uparrow} +\left(\epsilon^{\phantom{\dagger}}_{\k}+\Delta_0\right) \hat{c}^{\dagger}_{\k\downarrow}\hat{c}^{\phantom{\dagger}}_{\k\downarrow} + \xi^{\phantom{\dagger}}_{\k}\hat{f}^{\dagger}_{\k}\hat{f}^{\phantom{\dagger}}_{\k} \right],  \\
\hat{H}_{\Omega} &= \frac{\Omega_0}{2}\sum_{\k}\left( \hat{c}^{\dagger}_{\k\uparrow}\hat{c}^{\phantom{\dagger}}_{\k\downarrow} + \hat{c}^{\dagger}_{\k\downarrow}\hat{c}^{\phantom{\dagger}}_{\k\uparrow}  \right), \\
\hat{V} &= g\sum_{\k\k'\q} \hat{c}^{\dagger}_{\k\uparrow}\hat{c}^{\phantom{\dagger}}_{\k+\q\uparrow} \hat{f}^{\dagger}_{\k'}\hat{f}^{\phantom{\dagger}}_{\k'-\q}.\label{eq:HI} \end{alignat}
\end{subequations}
Here, $\hat H_0$, $\hat H_\Omega$, and $\hat{V}$ describe, respectively the kinetic energies of the particles, the Rabi drive, and the impurity-medium interactions. The operator $\hat{c}^{\dagger}_{\k\sigma}$ ($ \hat{c}_{\k\sigma}$) creates (destroys) an impurity with spin-$\sigma$, mass $m$, and momentum $\k$, where the corresponding 
kinetic energy is $\epsilon_\k=|\k|^2/2m\equiv k^2/2m$.  %
We assume that only the spin-$\up$ impurity interacts with the medium particles, where $g$ denotes the bare interaction strength, but it is straightforward to generalize our approach to the case where both spins interact with the medium~\cite{Adlong2020}. The interacting $\up$ impurity is coupled to the non-interacting $\down$ impurity via the Rabi coupling $\Omega_0$, where the Rabi drive is detuned from the bare $\up$-$\down$ transition by $\Delta_0$, and we have applied the rotating wave approximation. We emphasize that the behavior in the single-impurity limit also applies to a finite impurity density as long as we are in a regime where we can neglect correlations between impurities. At finite temperature, this is always true at sufficiently low impurity density.
Throughout, we work in units where the reduced Planck constant $\hbar$, the Boltzmann constant $k_B$, and the volume are all set to 1.

The fermionic operator $\hat{f}^\dag_\k$ ($\hat{f}_\k$) creates (annihilates) medium particles with momentum $\k$ and kinetic energy $\xi_\k = \epsilon_\k - \mu$. Here we assume that the fermionic medium particles have the same mass as the impurity, a scenario which is routinely accessed in $^6$Li atomic gases \cite{Scazza2017}, and we treat the Fermi gas within the grand canonical ensemble, such that it is described by a chemical potential $\mu$ and temperature $T$. We can determine the density $n$ of the Fermi medium using %
\begin{align}
n = \sum_\q %
n_\q 
= -\left(\frac{m T}{2\pi}\right)^{3/2}{\rm Li}_{3/2}\left( -e^{\beta\mu}   \right),
\end{align}
where the Fermi-Dirac distribution $n_\q = (1+e^{\beta \xi_\q})^{-1}$, $\beta \equiv 1/T$, and ${\rm Li}$ is the polylogarithm. We furthermore define the Fermi momentum and energy: $k_{\rm F} = (6\pi^2 n)^{1/3}$ and $\ef =k_{\rm F}^2/2m$, respectively, as well as the corresponding Fermi temperature $T_{\rm F}=\ef$ and time $\tau_{\rm F} = 1/\ef$. \replyadd{Throughout, we assume that the impurity particle experiences a uniform medium, as is the case for quantum gases in box traps~\cite{Mukherjee2017,Hueck2018} or for a low density of impurities in the center of a harmonically trapped gas.}

The properties of the driven impurity are encoded in the retarded impurity Green's function:
\begin{subequations}
\label{eq:Gt}
\begin{align}
 G_{\sigma}(\p,t) & = -i \theta(t)%
 \expval*{\hat{c}^{\phantom{\dagger}}_{\p\sigma}(t) \hat{c}^{\dagger}_{\p\sigma}(0)}, \\ %
 G_{\sigma\bar\sigma}(\p,t) & %
 = -i \theta(t)%
 \expval*{\hat{c}^{\phantom{\dagger}}_{\p\sigma}(t) \hat{c}^{\dagger}_{\p\bar\sigma}(0)}, %
\end{align}
\end{subequations}
with the spin components $\sigma \neq \bar\sigma$. Here, $\theta(t)$ is the Heaviside function, and we have used the time-dependent operator $\hat{c}^{\phantom{\dagger}}_{\p\sigma}(t) = e^{i\hat{H}t} \hat{c}^{\phantom{\dagger}}_{\p\sigma} e^{-i\hat{H}t}$. We define $\langle \cdots \rangle\equiv \Tr[e^{-\beta \hat{H}} \cdots ]_{\rm med}/\Tr[e^{-\beta \hat{H}}]_{\rm med}$, where the trace is over medium-only states since we are taking the single-impurity limit.

From the Fourier transform of the impurity Green's function, we can define the spin-resolved impurity spectral functions 
\begin{align}\label{eq:spectral}
    A_\sigma (\p,\omega) = -\frac{1}{\pi}{\rm Im}\, G_\sigma(\p,\omega+i0). 
\end{align}
Here, and in the following, we assume that the frequency in the Green's function is shifted infinitesimally into the upper half plane due to the unit step function in Eq.~\eqref{eq:Gt}. The Green's functions can also be formally linked to the thermodynamic properties in the long-time limit, as we discuss in Sec.~\ref{sec:thermodyn}.

\subsection{Non-interacting impurity Green's function}

\begin{figure}    \centering\includegraphics{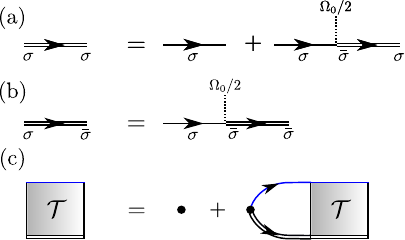}\centering
    \caption{\replyadd{(a,b) Relationships between the diagonal and off-diagonal Rabi-coupled Green's functions for spin $\sigma=\up,\down$ and $\bar\sigma\neq\sigma$. Thin solid lines represent the uncoupled Green's function, dotted lines the Rabi coupling between spin states, double lines represent the Rabi dressed Green's function. (c) Diagrammatic representation of the $T$-matrix equation, where the circle represents the interaction $g$, the thin blue line is the Green's function of a particle from the medium, and the double line is the Rabi-coupled Green's function $G_\up^{(\Omega)}$.
    }
    }
    \label{fig:Greens_diagram}
\end{figure}

\replyadd{In the absence of impurity-medium interactions and Rabi coupling, i.e., for $g=0$ and $\Omega_0=0$, the impurity Green's functions take the form
\begin{align}
    G_\up^{(0)}(\p,\omega)=\frac1{\omega-\ep}, \qquad 
    G_\down^{(0)}(\p,\omega)=\frac1{\omega-\ep-\Delta_0}.
\end{align}
Turning on Rabi coupling but keeping interactions off, the corresponding Green's function $\mathbf{G}_\Omega$ is then a matrix. In the $\up$-$\down$ basis we have
\begin{alignat}{1} \label{eq:GOmegamat}
\mathbf{G}_\Omega(\p,\omega)&= \fourvec{G_{\up}^{(\Omega)}}{G_{\up\down}^{(\Omega)}}{G_{\down\up}^{(\Omega)}}{G_{\down}^{(\Omega)}}  \nn \\
&=\fourvec{[G_\up^{(0)}(\p,\omega)]^{-1}}{-\Omega_0/2}{-\Omega_0/2}{[G_\down^{(0)}(\p,\omega)]^{-1}}^{-1}.
\end{alignat}
The diagonal and off-diagonal Green's functions satisfy the diagrammatic equations illustrated in Fig.~\ref{fig:Greens_diagram}(a) and (b).}

Inverting the matrix \replyadd{in Eq.~\eqref{eq:GOmegamat}}, we arrive at
\begin{subequations} \label{eq:G0}
\begin{alignat}{1}
G_{\up}^{(\Omega)}(\p,\omega) & = \frac{v^2}{\omega-E_{\p}^{+}} + \frac{u^2}{\omega-E_{\p}^{-}},
\label{eq:G0up}\\
G_{\up\down}^{(\Omega)}(\p,\omega) & = \frac{u v}{\omega-E_{\p}^{+}} - \frac{u v}{\omega-E_{\p}^{-}}, \\
G_{\down\up}^{(\Omega)}(\p,\omega) & = \frac{u v}{\omega-E_{\p}^{+}} - \frac{u v}{\omega-E_{\p}^{-} }, \\
G_{\down}^{(\Omega)}(\p,\omega) & = \frac{u^2}{\omega-E_{\p}^{+}} + \frac{v^2}{\omega-E_{\p}^{-}} .
\end{alignat}
\end{subequations}
\replyadd{The poles of the Rabi-coupled Green's functions yield}
the {Rabi-coupled} single-particle energies
\begin{align}\label{eq:Esingle}
    E_{\p}^{\pm} = \epsilon_\p+\frac12\left(\Delta_0 \pm \sqrt{\Omega_0^2 +\Delta_0^2} \right) ,
\end{align}
\replyadd{i.e., we have two Rabi-split parallel dispersions. These correspond to the stationary states in the absence of interactions. Furthermore, we have defined} the coefficients 
\begin{alignat}{1}
u^2 &= \frac{1}{2}\left(1 + \frac{\Delta_0}{\sqrt{\Omega_0^2 +\Delta_0^2} }  \right), \nonumber \\
v^2 &= \frac{1}{2}\left(1 - \frac{\Delta_0}{\sqrt{\Omega_0^2 +\Delta_0^2} }  \right), \nonumber \\ \label{eq:coeffs}
uv &= \frac{1}{2}\frac{\Omega_0}{\sqrt{\Omega_0^2 +\Delta_0^2} },
\end{alignat}
with $u^2+v^2=1$. \replyadd{Upon inspection of the diagonal elements of the Rabi-coupled Green's function in Eq.~\eqref{eq:G0}, we see that} $u^2$ and $v^2$ are the spin-$\up$ fractions of the lower and upper branches, respectively \replyadd{(and vice versa for the spin-$\down$ fractions).}

The corresponding spectral functions are obtained from Eq.~\eqref{eq:G0} and take the form
\begin{subequations}\label{eq:A0}
\begin{align}
    A^{(\Omega)}_\up(\p,\omega) & = v^2 \delta(\omega - E_\p^+) + u^2 \delta(\omega - E_\p^-), \\
    A^{(\Omega)}_\down(\p,\omega) & = u^2 \delta(\omega - E_\p^+) + v^2 \delta(\omega - E_\p^-).
\end{align}
\end{subequations}
Thus, in the absence of interactions we have two delta-function peaks. %

\subsection{Two-body problem}
We now turn to consider the interacting problem, focusing first on the two-body problem of a spin-$\up$ impurity interacting with a single medium particle. %
To model this, we consider a short-range contact interaction as in $\hat{V}$ in Eq.~\eqref{eq:HI}, whose bare interaction strength $g$ satisfies the renormalization condition:
\begin{alignat}{1}
\frac{1}{g} = \frac{m}{4\pi a_{\rm s}} - %
\sum_{\k}^{\Lambda} \frac{1}{2\epsilon_\k}. 
\label{eq:ginv}
\end{alignat}
Here, $a_{\rm s}$ is the scattering length while $\Lambda$ is an ultraviolet (UV) cutoff on the relative momentum in the scattering process. In all results presented in this work, we have taken the UV cutoff to infinity.

In the absence of Rabi coupling, the two-body problem features a bound state for $a_{\rm{s}}>0$, with binding energy $\eb=1/ma_{\rm{s}}^2$. This shows up as an energy pole of the two-body $T$ matrix \replyadd{describing the scattering of a spin $\up$ particle with a particle from the medium~\cite{parish2023fermi}:}
\begin{align}
    {\mathcal T}^{-1}_0(\omega) &= 
    \frac{1}{g} \replyadd{-\sum_{\k}^{\Lambda} %
    G_\up^{(0)}(\k,\omega-\ek)}
\nn    \\ &
    =
    \frac{m}{4\pi \as} - \frac{m^{3/2}}{4\pi}\sqrt{-\omega} .
\label{eq:TvacnoRabi}
\end{align}
Here we have used Eq.~\eqref{eq:ginv} and taken the limit $\Lambda\to\infty$ in the second step.

In the presence of Rabi coupling, the two-body $T$ matrix \replyadd{between a spin $\up$ particle and a particle from the medium} is modified to
\begin{align}
    {\mathcal T}^{-1}_\Omega(\omega)&= 
    \frac{1}{g} %
    \replyadd{-\sum_{\k}^{\Lambda} %
    G_\up^{(\Omega)}(\k,\omega-\ek)}\nn \\ &=\frac{m}{4\pi \as} - \frac{m^{3/2}}{4\pi}\left(u^2\sqrt{E_{\mathbf{0}}^--\omega}+v^2\sqrt{E_{\mathbf{0}}^+-\omega} \right),
\label{eq:TvacRabi}
\end{align}
\replyadd{as illustrated in Fig.~\ref{fig:Greens_diagram}(c). Importantly, all intermediate free propagation involves the Rabi-coupled spin-$\up$ Green's function in Eq.~\eqref{eq:G0up}.} %
We see that the continuum instead starts at $E^-_\0$, and that there is only a two-body bound state \replyadd{(corresponding to a stationary state of the two-body problem)} when $1/\as>1/a_\mathrm{c}$ with 
\begin{align}\label{eq:acrit}
    1/a_\mathrm{c}\equiv v^2\sqrt{m(E_\0^+-E_\0^-)}\geq0.
\end{align}
For this range of scattering lengths, the $T$ matrix has a pole at
\begin{align}
\label{eq:2bodyomega}
    \omega=-\frac1{m\as^2}\left[1+\frac{\Omega_0^2}{4\Delta_0^2}\left(\sqrt{1+m\as^2\Delta_0}-1\right)^2\right].
\end{align}
This is always below $-1/m\as^2$; however note that the actual binding energy is reduced compared with the $\up$ impurity in the absence of a Rabi drive.

\begin{figure}
    \centering
    \includegraphics{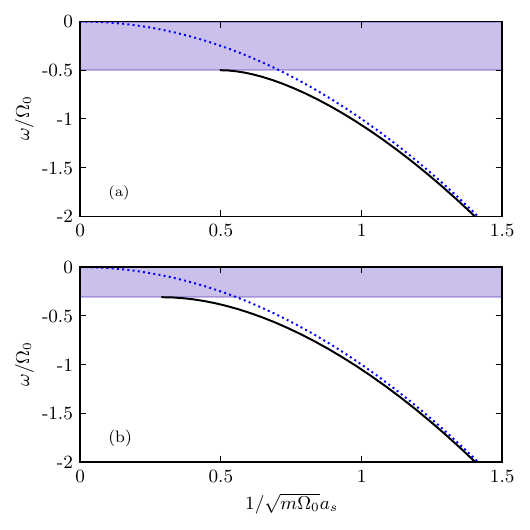}\centering
    \caption{Energy of the two-body bound state (black, solid line) for detuning (a) $\Delta_0=0$, and (b) $\Delta_0=0.5\Omega_0$. We also plot the energy in the absence of Rabi coupling (blue dotted). The  continuum is shown as a purple shaded region.
    }
    \label{fig:2body}
\end{figure}

Figure~\ref{fig:2body} demonstrates the shift of the continuum and the associated shift of the critical scattering length at which the two particles can bind. We see that the energy of the bound state is only strongly modified in the vicinity of the critical scattering length, while it is insensitive to the Rabi drive when the binding energy becomes large compared with the drive parameters.

\replyadd{To round off our discussion of two-body physics, we note that we can also calculate the scattering lengths corresponding to scattering of one of the Rabi-coupled single-particle states with a medium particle. In either case, the scattering length involves the fraction of spin-$\up$ particle in the Rabi-coupled state, and the $T$ matrix should be evaluated at the appropriately shifted collision energy. For instance, the scattering length for the lower Rabi-coupled state is
\begin{align}
    a_- &=u^2\frac{m}{4\pi}\mathcal{T}_\Omega(E^-_\0) \nn \\
    &=u^2 \frac1{a_\mathrm{s}^{-1}-a_\mathrm{c}^{-1}}.\label{eq:aminus}
\end{align}
We see that this diverges at the critical scattering length for bound-state formation, $a_\mathrm{c}$, as one would expect. Equation~\eqref{eq:aminus} makes it clear that our scenario is qualitatively distinct from earlier proposals to modify the two-body physics via an oscillating magnetic field~\cite{Smith2015,Sykes2017}, since in that case the scattering length always contains an imaginary part.

Similarly to Eq.~\eqref{eq:aminus}, we can define the scattering length for the upper Rabi-coupled state, $a_+=v^2 (m/4\pi)\mathcal{T}_\Omega(E^+_\0)$. This is complex since the upper Rabi-coupled state is not the single-particle ground state. Similarly, there is not a true bound state of the upper Rabi-coupled state and a medium particle (corresponding to a pole of $\mathcal{T}_\Omega(\omega)$ for real $\omega$), only a quasibound state (a pole off the real axis) as expected when Rabi-coupling to a continuum of scattering states~\cite{LiFloquet1999}.
}

\begin{figure}
    \centering\includegraphics{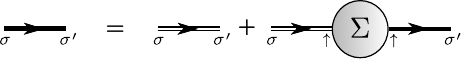}\centering
    \caption{\replyadd{Dressed interacting Green's function (solid line) for spin states $\sigma=\up,\down$. Double lines represent the Rabi dressed Green's function. The self-energy is shown as a circle, and it only couples spin-$\up$ particles, since we assume that the spin-$\down$ state is non-interacting.
    }
    }
    \label{fig:Dyson_diagram}
\end{figure}

\subsection{Interacting Green's function}

The presence of impurity-medium interactions leads to a modified impurity Green's function. This can be captured by introducing the interaction-induced self-energy $\boldsymbol{\Sigma}$. \replyadd{The resulting Dyson equation is shown in Fig.~\ref{fig:Dyson_diagram}, which can be written as follows}~\cite{fetterbook}:
\begin{alignat}{1} \label{eq:green}
\mathbf{G}(\p,\omega) =  \left( \mathbf{G}^{-1}_{\Omega}(\p,\omega) - \mathbf{\Sigma}(\p,\omega)  \right)^{-1}.
\end{alignat}
Since we assume that there are only interactions between the $\up$ impurity and the medium, we have 
\begin{alignat}{1}
\mathbf{\Sigma}(\p,\omega)  =  \fourvec{\Sigma(\p,\omega)}{0}{0}{0}.
\end{alignat}
This can straightforwardly be modified to include interactions between the spin-$\down$ impurity and the medium~\replyadd{\cite{Adlong2020,hu2022Rabi}}. Importantly, the self-energy depends itself on the Rabi drive, since the impurity can change its spin inside the diagrams contributing to the self-energy.

The diagonal parts of the impurity Green's function, $ G_{\up} \equiv \mathbf{G}_{11} $ and $  G_{\down} \equiv \mathbf{G}_{22}$, are now given by
\begin{subequations} \label{eq:G}
\begin{alignat}{1}
G_{\uparrow}(\p,\omega) &= \frac{1}{\omega-\epsilon_\p-\Sigma(\p,\omega)-\left(\frac{\Omega_0}{2}\right)^2\frac{1}{\omega-\epsilon_\p-\Delta_0}} ,\\
G_{\downarrow}(\p,\omega) &= \frac{1}{\omega-\epsilon_\p-\Delta_0 -\left(\frac{\Omega_0}{2}\right)^2\frac{1}{\omega-\epsilon_\p-\Sigma(\p,\omega)} }. \label{eq:G_down}
\end{alignat}
\end{subequations}
From the Green's functions, we can obtain the quasiparticle energies using the usual relation 
${\rm{Re}}[G^{-1}_\sigma(\0,\mathcal{E}_{\sigma})]=0$,
which yields
\begin{subequations}\label{eq:epol}
\begin{align}
    \mathcal{E}_{\up} & = \Re\left[\Sigma(\0,\mathcal{E}_{\up})\right] + \left(\frac{\Omega_0}{2}\right)^2\frac{1}{\mathcal{E}_{\up}-\Delta_0}, \label{eq:epolup}\\
    \mathcal{E}_{\down} & = \Delta_0 + \left(\frac{\Omega_0}{2}\right)^2\Re\left[\frac{1}{\mathcal{E}_{\down}-\Sigma(\0,\mathcal{E}_{\down})}\right] .\label{eq:epoldown}
\end{align}
\end{subequations}
These quasiparticle energies are in general different, being identical only in the case of the polaron ground state at $T=0$ where the self-energy is purely real at the quasiparticle pole. Note that with increasing temperature, the polaron quasiparticle may cease to be well defined in the sense that one or both expressions in Eq.~\eqref{eq:epol} have no solution.

Calculating the impurity self-energy is in general a complicated many-body problem. In Sec.~\ref{sec:approx} we will discuss how this can be obtained within a many-body $T$-matrix approximation that is equivalent to a variational approach~\cite{Parish2016,Liu2019}. However, as we now discuss in Sec.~\ref{sec:thermodyn}, we can make several observations about the impurity thermodynamics that are independent of the precise form of the self-energy.

\section{Exact thermodynamic properties} \label{sec:thermodyn}

We can investigate the thermodynamic properties of the Rabi-coupled impurity in the long-time limit, where the interacting impurity has fully thermalized with the medium. In this case, we can define an impurity free energy $\mathcal{F} = F - F_0$, where $F$ and $F_0$ correspond, respectively, to the free energies of the Rabi-coupled interacting system and of the non-interacting system without Rabi coupling. We define it in this way in order to remove the extensive properties of the medium and to isolate the non-trivial behavior of the Rabi-driven impurity. 

Within the grand canonical ensemble for the medium, the free energy $\mathcal{F}$ depends on the medium chemical potential $\mu$, the temperature $T$ and the medium-impurity scattering length $\as$, as well as the Rabi-drive parameters $\Delta_0$ and $\Omega_0$. By taking derivatives with respect to these quantities, we obtain key observables that characterize the impurity thermodynamics. In particular, the magnetization is %
\begin{align}\label{eq:Sz}
    \mathcal{S}_z & \equiv N_{\uparrow}-N_{\downarrow} =1 - 2 \frac{\partial \mathcal{F}}{\partial\Delta_0} ,
\end{align}
where $N_\sigma = \sum_\k \Tr[\hat\rho \, \hat{c}^{\dagger}_{\k\sigma} \hat{c}_{\k\sigma} ]$ and we have used the fact that we are considering a single impurity: $N_\up + N_\down = 1$. The trace is over all states containing the medium \textit{and} the single impurity (note the difference from the angle brackets defined previously), and we define the corresponding density matrix $\hat\rho\equiv e^{-\beta \hat{H}}/\Tr[e^{-\beta \hat{H}}]$. Similarly to $\mathcal{S}_z$, we have the in-plane spin/magnetization along the $x$-direction:
\begin{align} \label{eq:Sx}
    \mathcal{S}_x & \equiv \sum_\k \Tr[\hat\rho \,\left(\hat{c}^{\dagger}_{\k\up} \hat{c}_{\k\down} + \hat{c}^{\dagger}_{\k\down} \hat{c}_{\k\up}\right) ] = 2 \frac{\partial \mathcal{F}}{\partial\Omega_0} .
\end{align}
There are also additional quantities that directly capture the correlations due to interactions with the medium. Specifically, we have the Tan contact \cite{Tan1}
\begin{align} \label{eq:cont_def}
\mathcal{C} = 4\pi m \frac{\partial \mathcal{F}}{\partial \left( -1/\as\right) },
\end{align}
and the number of medium atoms in the dressing cloud of the impurity \cite{Massignan2005}
\begin{align} \label{eq:N}
\mathcal{N} =  -\frac{\partial \mathcal{F}}{\partial \mu}.
\end{align} 

In the absence of impurity-medium interactions, the free energy takes the simple form for a spin-$1/2$ system
\begin{align} \label{eq:spinF}
    \mathcal{F} = \frac{\Delta_0}{2} - \frac{1}{\beta}\ln\left[2 \cosh(\frac{\beta}{2}\sqrt{\Omega_0^2 + \Delta_0^2}) \right] ,
\end{align}
which yields the magnetizations
\begin{subequations}
\begin{align}
    \mathcal{S}_z & = \frac{\Delta_0}{\sqrt{\Omega_0^2 + \Delta_0^2}} \tanh(\frac{\beta}{2}\sqrt{\Omega_0^2 + \Delta_0^2}), \\ 
    \mathcal{S}_x & = -\frac{\Omega_0}{\sqrt{\Omega_0^2 + \Delta_0^2}} \tanh(\frac{\beta}{2}\sqrt{\Omega_0^2 + \Delta_0^2}) .   
\end{align}
\end{subequations}
In the zero-temperature limit, we recover the ground-state energy $\mathcal{F} = E_\0^-$ from Eq.~\eqref{eq:Esingle} with corresponding magnetizations $\mathcal{S}_z =\Delta_0/\sqrt{\Delta_0^2+\Omega_0^2}= u^2 - v^2$ and $\mathcal{S}_x  =-\Omega_0/\sqrt{\Delta_0^2+\Omega_0^2}=-2uv$ in terms of the coefficients defined in Eq.~\eqref{eq:coeffs}.

In the general case, using dimensional analysis~\cite{Braaten2008,werner2012virial}, we can write the free energy in terms of an arbitrary length scale $\lambda$ as follows: $\mathcal{F}(T,\mu,\as,\Delta_0,\Omega_0) \!  = \lambda^{-2} \mathcal{F}(T\lambda^2,\mu\lambda^2,\as/\lambda,\Delta_0\lambda^2,\Omega_0\lambda^2)$. Taking $d\mathcal{F}/d\lambda=0$ and then setting $\lambda=1$, we obtain 
\begin{align} \notag
\mathcal{F} = \left(T \partial_T + \mu\partial_\mu+ \frac{1}{2\as}\partial_{1/\as} + \Delta_0\partial_{\Delta_0}  + \Omega_0\partial_{\Omega_0}  \right) \mathcal{F} ,
\end{align}
which gives us a relationship between all the thermodynamic properties of the impurity,
\begin{align}
  \mathcal{F} = - T S   - \mu \mathcal{N} - \frac{\mathcal{C}}{8 \pi m \as} + \frac{\Delta_0}{2}(1-\mathcal{S}_z)  + \! \frac{\Omega_0}{2} \mathcal{S}_x ,
\end{align}
where the impurity entropy $S = - \partial_T \mathcal{F}$. In particular, for a unitary Fermi gas in the zero-temperature limit, the impurity free energy is solely related to the number of atoms in the dressing cloud and the magnetizations:
\begin{align}
\label{eq:freeenergyunitarityt0}
    \mathcal{F}_{1/a=0,T=0} =  - \ef  \mathcal{N} + \frac{\Delta_0}{2}(1-\mathcal{S}_z)  + \! \frac{\Omega_0}{2} \mathcal{S}_x .
\end{align}

In practice, we can derive the free energy from the full spectrum contained in the Green's function \eqref{eq:green}. Following the arguments in Refs.~\cite{Liu2020PRA,Liu2020PRL} (see Appendix~\ref{app:freeF} for details), the impurity free energy satisfies the relation 
\begin{alignat}{1} \label{eq:freeenergy_rel}
e^{-\beta\mathcal{F}} = \frac{ \int d\omega\sum_{\mathbf{p}}e^{-\beta\omega}\left[ A_{\up}(\p,\omega) + A_{\down}(\p,\omega)  \right]}{\sum_{\mathbf{p}} e^{-\beta\epsilon_{\mathbf{p}}}},
\end{alignat}
in terms of the spin-resolved impurity spectral functions. Using the spectral functions in the absence of interactions, Eq.~\eqref{eq:A0}, it is straightforward to show that we recover the simple spin-$1/2$ case in Eq.~\eqref{eq:spinF} when the Green's function is non-interacting. 

The impurity free energy becomes particularly simple when there is a well-defined polaron quasiparticle in the limits $T\ll\ef$ and $T\ll\Omega_0$, such that we can assume that the imaginary part of the impurity self-energy is negligible at the quasiparticle pole. The quasiparticle energies then coincide, $\mathcal{E}_\up=\mathcal{E}_\down\equiv\mathcal{E}$, where
\begin{align} \label{eq:selfconsistent}
\mathcal{E} = \left( \frac{\Omega_0}{2} \right)^2 \frac{1}{\mathcal{E}-\Delta_0} + \Sigma(\0,\mathcal{E}).
\end{align}
This simplification allows us to gain analytic insights in the weak and strong Rabi-drive limits, as we now discuss.

\subsection{Weak Rabi drive}\label{sec:weak}

It is instructive to first consider the case of a small Rabi coupling, $\Omega_0 \ll \ef $, where the steady-state properties are only weakly perturbed by the rf drive. From Eq.~\eqref{eq:selfconsistent}, the lowest order term where $\Omega_0 =0$ yields the attractive $\up$ polaron energy in the absence of Rabi coupling
\begin{align} \label{eq:polE} 
\mathcal{E}_0 = \Sigma(\0,\mathcal{E}_0)|_{\Omega_0 = 0} \equiv \Sigma_0(\mathcal{E}_0) .
\end{align}
Now we wish to perform an expansion in small $\Omega_0/\ef $ for the energy $\mathcal{E}$ and self-energy,
\begin{align}
    \mathcal{E} & = \mathcal{E}_0 + \delta \mathcal{E}^{(1)} %
    + \cdots \\
    \Sigma(\0,\mathcal{E}) & = \Sigma_0(\mathcal{E}_0)  + \delta \Sigma^{(1)} (\mathcal{E}_0) + \left.\frac{\partial \Sigma_0}{\partial \mathcal{E}}\right|_{\mathcal{E}_0} \! \delta \mathcal{E}^{(1)} + 
    \cdots 
\end{align}
The linear order contribution $\delta \Sigma^{(1)} (\mathcal{E}_0) = 0$ since, on physical grounds, the self-energy must be an even function of the Rabi coupling $\Omega_0$. Moreover, the self-energy is expected to be analytic around $\Omega_0 = 0$ provided we are away from the bare resonance $\Delta_0 = 0$. Therefore, keeping only linear terms and using Eq.~\eqref{eq:polE}, Eq.~\eqref{eq:selfconsistent} becomes
\begin{align} \label{eq:smallR}
\underbrace{\left(1 - \left.\frac{\partial \Sigma_0}{\partial \mathcal{E}}\right|_{\mathcal{E}_0}\right)}_{Z_0^{-1}} \delta \mathcal{E}^{(1)} \simeq \left(\frac{\Omega_0}{2}\right)^2 \frac{1}{\delta \mathcal{E}^{(1)} - \delta} ,
\end{align}
where $Z_0$ is the residue of the attractive $\up$ polaron in the absence of Rabi coupling, and we have defined the detuning from the $\up$ polaron energy:  $\delta = \Delta_0 - \mathcal{E}_0$. Note that  the right hand side of Eq.~\eqref{eq:smallR} is also effectively linear in $|\Omega_0|$ if $|\delta| < |\Omega_0| \ll \ef $. 

Solving for $\delta \mathcal{E}^{(1)}$ then yields the ground and excited polaron energies of the Rabi coupled system:
\begin{align} \label{eq:pol_ener_app}
\mathcal{E}^{\pm} = \mathcal{E}_0 +  \frac{1}{2}\left(\delta \pm  \sqrt{\delta^2 + \Omega^2} \right).
\end{align}
Notice how this resembles Eq.~\eqref{eq:Esingle}, the spin-$1/2$ case in the absence of a medium, where $\Omega \equiv \sqrt{Z_0} \Omega_0$ is the effective Rabi frequency for the polaron quasiparticle, consistent with previous work~\cite{Kohstall2012,Adlong2020}. 

Since we have assumed that $T \ll \ef $, the temperature is sufficiently low that we can use these quasiparticle energies to derive a free energy in a similar manner to the spin-$1/2$ case in the absence of a medium. Using Eq.~\eqref{eq:freeenergy_rel}, we obtain
\begin{align} \label{eq:free_weak}
    \mathcal{F} =  \mathcal{E}_0 +  \frac{\delta}{2} - \frac{1}{\beta}\ln\left[2 \cosh(\frac{\beta}{2}\sqrt{\Omega^2 + \delta^2}) \right] .
\end{align}
From Eqs.~\eqref{eq:Sz} and \eqref{eq:Sx}, this gives the corresponding magnetizations
\begin{subequations}
\label{eq:Sweak}
\begin{align} 
\label{eq:Szweak}
    \mathcal{S}_z & = \frac{\delta}{\sqrt{\delta^2 + \Omega^2}} \tanh\left(\frac{\beta}2 \sqrt{\delta^2 + \Omega^2}\right), \\
\label{eq:Sxweak}
    \mathcal{S}_x & = -\frac{\sqrt{Z_0}\Omega}{\sqrt{\delta^2 + \Omega^2}} \tanh\left(\frac{\beta}2 \sqrt{\delta^2 + \Omega^2}\right),
\end{align}
\end{subequations}
which, in the zero-temperature limit, reduce to
\begin{subequations}
\label{eq:Sweak0}
\begin{align} 
\label{eq:Szweak0}
    \mathcal{S}_z & = \frac{\delta}{\sqrt{\delta^2 + \Omega^2}}, \\
\label{eq:Sxweak0}
    \mathcal{S}_x & = -\frac{\sqrt{Z_0}\Omega}{\sqrt{\delta^2 + \Omega^2}} .
\end{align}
\end{subequations}

From Eqs.~\eqref{eq:Szweak} and \eqref{eq:Szweak0}, 
we find that we always have $\mathcal{S}_z = 0$ when the Rabi drive is resonant with the $\up$ polaron energy, $\Delta_0 = \mathcal{E}_0$, a result which only depends on temperature via $\mathcal{E}_0$~\footnote{At finite temperature, we determine the polaron quasiparticle properties in the absence of Rabi coupling  in the usual way, where the polaron energy $\mathcal{E}_0 = \Re\Sigma(\0,\mathcal{E}_0)$, the residue $Z_0^{-1} = 1- \partial_{\mathcal{E}}\Re\left.\Sigma(\0,\mathcal{E})\right|_{\mathcal{E}_0}$, and the damping rate $\Gamma_0 = - Z_0 \Im \Sigma(\0,\mathcal{E}_0)$ \label{foot:norabi} }. 
However, the total magnetization is sensitive to both temperature and $\Omega$, as we can see from the slope around the resonance
\begin{align}
\left. \frac{\partial \mathcal{S}_z}{\partial \Delta_0} \right|_{\Delta_0 = \mathcal{E}_0} = \frac{\tanh(\beta\Omega/2)}{\Omega}.
\end{align}
Furthermore, close to the zero crossing of $\mathcal{S}_z$, Eq.~\eqref{eq:Sxweak0} yields $\mathcal{S}_x=-\sqrt{Z_0}$ rather than $-1$ at zero temperature, thus illustrating how interactions modify the impurity spin.

\subsection{Strong Rabi drive}\label{sec:strong}

In the opposite limit of a strong Rabi drive relative to all other parameters, the impurity approaches a freely driven two-level system since the time-scale of many-body excitations, $1/\ef$, greatly exceeds the period of the drive. We can then calculate the leading-order many-body corrections as follows. 

\replyadd{To expose the behavior at large Rabi coupling, we assume that $\Omega_0$ greatly exceeds $|\Delta_0|$ and $E_\mathrm{F}$, and  furthermore, if $a_\mathrm{s}>0$, that $\Omega_0\gg \varepsilon_\mathrm{B}$. This ensures that the polaron energy in the driven system is dominated by the single-particle energy $\sim -\Omega_0/2$ and not by many-body effects of $O(E_{\mathrm{F}})$ or effects related to the two-body bound state. To leading order in $E_\mathrm{F}/\Omega_0$}
we can therefore approximate $\mathcal{E}\simeq E_\0^-$ in the self-energy in Eq.~\eqref{eq:selfconsistent} \replyadd{(note that a similar approximation was used in Ref.~\cite{Shkedrov2022})}. Furthermore, in the large-drive limit, multiple excitations of the medium become suppressed, and we can consider the interaction of the impurity with only a single excitation of the medium. Thus, we can approximate the self-energy as arising due to the vacuum $T$ matrix in Eq.~\eqref{eq:TvacRabi}:
\begin{align}    \Sigma(\0,\mathcal{E})&\simeq n\mathcal{T}_{\Omega}(E_\0^-) 
    = \frac{4\pi n}m\left(
    \frac1{\as}-\frac1{a_\mathrm{c}}\right)^{-1},
    \label{eq:Sigmalarge}
\end{align}
with $a_\mathrm{c}$ the critical scattering length for forming a two-body bound state, as defined in Eq.~\eqref{eq:acrit}. The polaron energy then takes the form
\begin{align}
\mathcal{E}=\frac12\left[\Delta_0+\Sigma(\0,E_\0^-)-\sqrt{\left(\Delta_0-\Sigma(\0,E_\0^-)\right)^2+\Omega_0^2}\right].
\end{align}
In the strongly driven system, all other states with significant spectral weight are far detuned from the polaron energy, and hence we identify this with the impurity free energy. The various thermodynamic variables can then be obtained using Eqs.~\eqref{eq:Sz}-\eqref{eq:N}. 

In the special case of $1/\as=0$, we find the many-body corrections to the thermodynamic variables
\begin{subequations}
\begin{align}\label{eq:E-strong}
    \mathcal{E}&= -\frac{\Omega_0}2+\frac{\Delta_0}2-\frac{4\pi n}{m^{3/2}\sqrt{\Omega_0}},\\ \label{eq:Delta-c-large}
    \mathcal{S}_z&=\frac{\Delta_0}{\Omega_0}+\frac{16\pi n}{(m\Omega_0)^{3/2}},\\
    \mathcal{S}_x&=-1+\frac{4\pi n}{(m\Omega_0)^{3/2}}+\frac{\left(\Delta_0+\frac{16\pi n}{m^{3/2}\Omega_0^{1/2}}\right)^2}{2\Omega_0^2}, \\
    \mathcal{C}&=\frac{32 \pi^2n}{m \Omega_0}\left(
    1+3\frac{\Delta_0}{\Omega_0}\right)
    ,\\
    \mathcal{N}&=\sqrt{\frac{8\ef }{\pi^2\Omega_0}}\left(
    1+2\frac{\Delta_0}{\Omega_0}\right)
    ,
\end{align}
\end{subequations}
where we implicitly take $\Omega_0>0$.
Here we have assumed that we are close to the regime of zero magnetization along the $z$-axis, which is achieved when $\Delta_0=-16\pi n/(m^{3/2}\Omega_0^{1/2})$. In fact, these expressions are also valid away from unitarity, provided the Rabi coupling is sufficiently large such that the scattering length can be ignored in Eq.~\eqref{eq:Sigmalarge}.

\begin{figure}
    \centering\includegraphics{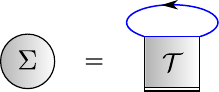}\centering
    \caption{\replyadd{The self-energy within the ladder approximation, with symbols defined as in Figs.~\ref{fig:Greens_diagram} and \ref{fig:Dyson_diagram}. %
    }
    }
    \label{fig:self_diagram}
\end{figure}

\section{Energy spectra and thermodynamics}
\label{sec:approx}

We now turn to the scenario of a Rabi drive of intermediate strength. In this case, we do not have any small parameters, and hence we must approximate the driven many-body problem. In this work, we utilize a many-body $T$-matrix approximation, which has been successfully applied to the Fermi polaron problem in the absence of Rabi coupling~\cite{combescot2007,Massignan2008PRA,Punk2009,bruun2010}. Our approach is equivalent to the result of the variational approach to Rabi dynamics introduced in Refs.~\cite{Parish2016,Adlong2020}, extended to finite temperature and impurity momentum using the formalism of Ref.~\cite{Liu2019}. Here, one truncates the time evolution of the impurity at the level of one excitation of the medium, and hence the self-energy can be approximated as a sum of ladder diagrams, \replyadd{shown in Fig.~\ref{fig:self_diagram}:}
\begin{equation}
    \Sigma(\p,\omega) = \sum_{\q} n_{\q} \mathcal{T}(\q+\p, \omega+\epsilon_{\q})\; ,
\label{eq:self-energy}
\end{equation}
where $n_\q$ is the Fermi-Dirac distribution for the medium. The in-medium $T$ matrix is \replyadd{obtained by taking Pauli blocking into account in all intermediate states, yielding}
\begin{align}  
    &\mathcal{T}^{-1}_{}(\q, \omega) \nn \\ 
& \quad = \frac{1}{g} \replyadd{-\sum_{\k}^{\Lambda} (1-n_\k)G_\up^{(\Omega)}(\q-\k,\omega-\ek)} \nn \\ 
    & \quad = \mathcal{T}_{\Omega}^{-1}\left(\omega-\frac{\epsilon_\q}{2}\right) \nn \\ &\qquad
    -\sum_{\k}n_\k\left(  \frac{ u^2}{\epsilon_{\k}+E_{\q-\k}^{-}-\omega}+\frac{ v^2}{\epsilon_{\k}+E_{\q-\k}^{+}-\omega} \right),\label{eq:invT-matrix} 
\end{align}
which depends on the Rabi coupling, similarly to the vacuum $T$ matrix in Eq.~\eqref{eq:TvacRabi}.
\begin{figure}[b]
    \centering
    \includegraphics{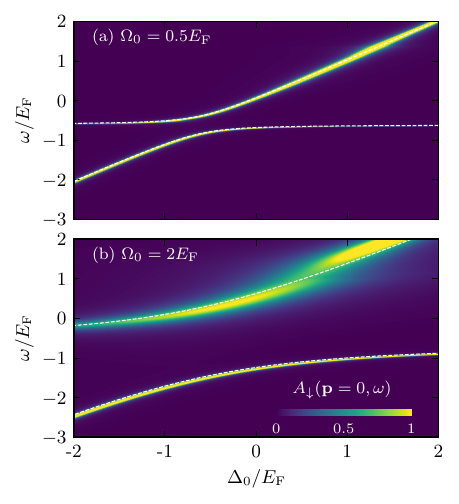}\centering
    \caption{ \replyadd{Color plot of the} zero-momentum impurity spectral function $A_{\downarrow}(\p=0,\omega)$ as a function of the detuning $\Delta_0/\ef $ for Rabi coupling (a) $\Omega_0=0.5\ef $ and (b) $\Omega_0=2.0\ef $. We take $1/\kf \as=0$, $T=0$, and for all spectra, a broadening of $0.005\ef $ has been added to enhance visibility. The dotted white lines are the polaron energies obtained from Eq.~\eqref{eq:pol_ener_app}. 
    }
    \label{fig:comp_ener}
\end{figure}

\begin{figure*}
    \centering
    \includegraphics{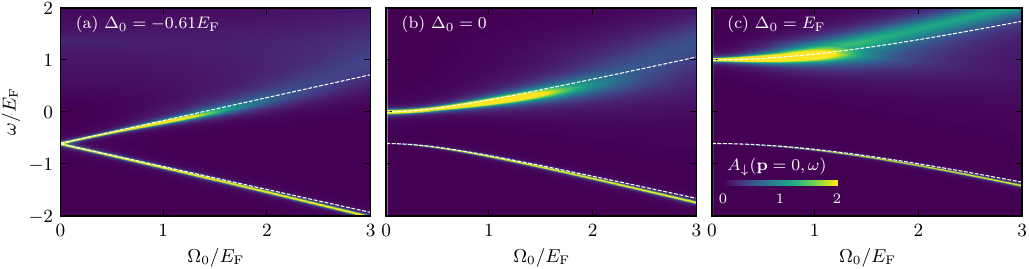}
    \caption{\replyadd{Color plot of the} zero-momentum impurity spectral functions $A_{\downarrow}(\p=0,\omega)$ as a function of the Rabi coupling $\Omega_0/\ef $ for detunings (a) $\Delta_0=E_{\rm att}=-0.61\ef$ (the attractive polaron energy in the absence of Rabi drive), (b) $\Delta_0=0$, and (c) $\Delta_0=\ef $. We take $1/\kf\as=0$, $T=0$, and for all spectra a broadening of $0.005\ef $ has been added to enhance visibility. The dashed white lines are the polaron energies obtained from Eq.~\eqref{eq:pol_ener_app}. 
    }
    \label{fig:comp}
\end{figure*}

\begin{figure*}
    \centering
    \includegraphics{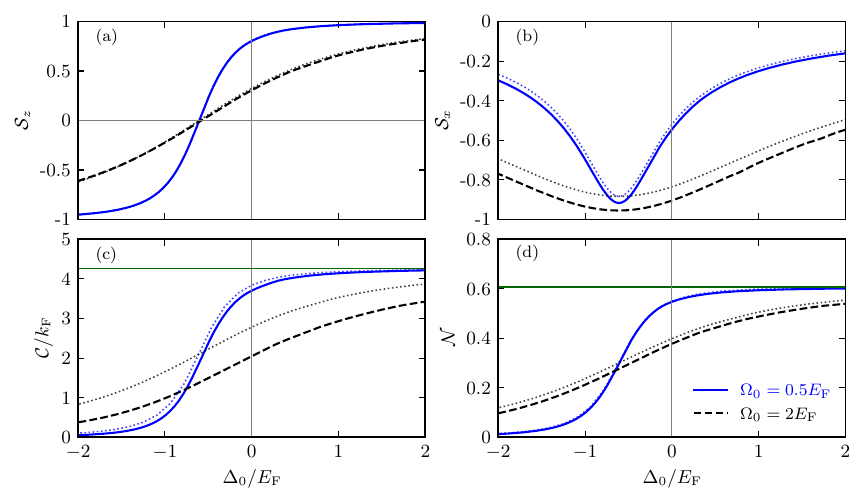}
    \caption{\replyadd{The magnetization $\mathcal{S}_z$  (a),  in-plane magnetization $\mathcal{S}_x$ (b), contact (c) and number of atoms in the dressing cloud  (d)} are plotted as a function of detuning $\Delta_0/\ef $ at zero temperature, $T=0$, interaction $1/\kf\as=0$, and for Rabi couplings $\Omega_0=0.5\ef $ (blue solid) and $\Omega_0=2.0\ef $ (black dashed). The dotted lines correspond to the weak Rabi drive in Eqs.~\eqref{eq:Sweak0}, \eqref{eq:Nweak0}, and \eqref{eq:Cweak0}, respectively,  and the green lines indicate the corresponding results in the absence of Rabi coupling. %
    }
    \label{fig:comp_T0}
\end{figure*}

Using the approximate self-energy in Eq.~\eqref{eq:self-energy}, we can calculate the impurity Green's functions in Eq.~\eqref{eq:G} and hence the impurity spectral functions in Eq.~\eqref{eq:spectral}. The resulting spectral function $A_\down(\0,\omega)$ at $T=0$ is shown in Figs.~\ref{fig:comp_ener} and \ref{fig:comp}. To be concrete, here and in the rest of this section we have taken the $\up$ impurity-medium interactions to be unitarity limited, $1/\kf\as=0$. This means that, within the $T$-matrix formalism and in the absence of a Rabi drive, the attractive polaron has energy $\mathcal{E}_0 \simeq-0.61\ef$~\cite{Chevy2006}, while the repulsive branch is a broad peak around the energy $1.5\ef$~\cite{Cui2010,Massignan2011}. Figure~\ref{fig:comp_ener} investigates the effect of varying the detuning at fixed Rabi drive. We see that when the Rabi drive $\Omega_0<\ef$, the peaks in the spectrum closely follow the quasiparticle energies obtained from solving the weak-drive expression in Eq.~\eqref{eq:pol_ener_app}, where the attractive polaron residue is $Z_0\simeq0.78$. On the other hand, for a larger Rabi drive where $\Omega_0>\ef$, we are clearly able to distinguish the broad continuum related to the coupling to the repulsive polaron, which leads to an effective avoided crossing between the repulsive branch and the upper energy in Eq.~\eqref{eq:pol_ener_app}. These features are reflected in Fig.~\ref{fig:comp}, where we instead fix the detuning and vary $\Omega_0$. In particular, we see that the additional features arising from the repulsive branch in the Rabi-coupled $\down$ spectral function require both a positive detuning and a large Rabi drive $\sim\ef$.

The impurity spectral functions in turn enable us to calculate the impurity free energy via Eq.~\eqref{eq:freeenergy_rel}, from which the thermodynamic variables in Eqs.~\eqref{eq:Sz}-\eqref{eq:N} are obtained by taking the appropriate derivatives. The results are shown in Fig.~\ref{fig:comp_T0} for zero temperature as a function of detuning and for the same Rabi couplings as for the spectra in Fig.~\ref{fig:comp_ener}. At zero temperature, the thermodynamic properties are entirely determined by the properties of the ground state. According to Eq.~\eqref{eq:Szweak0}, this is mostly spin $\down$ when $\Delta_0\lesssim -0.61\ef$, becoming increasingly dominated by the interacting spin $\up$ component with increasing detuning. Indeed, we see that this expression provides a near perfect agreement for the behavior of the magnetization $\mathcal{S}_z$, even for $\Omega_0=2\ef$.  %
We also observe that the position of the zero crossing of  $\mathcal{S}_z$ remains relatively %
constant as the Rabi drive is increased, with the corresponding critical detuning $\Delta_0^{(\mathrm{c})}$ being close to $-0.61\ef$ for both values of $\Omega_0$ shown. Thus, the weak-Rabi-drive result in Eq.~\eqref{eq:Szweak0} appears to be remarkably robust. 
Likewise, Eq.~\eqref{eq:Sxweak0} provides an excellent approximation of $\mathcal{S}_x$, which is maximal close to the zero crossing of $\mathcal{S}_z$.

\begin{figure*}
    \centering    
    \includegraphics{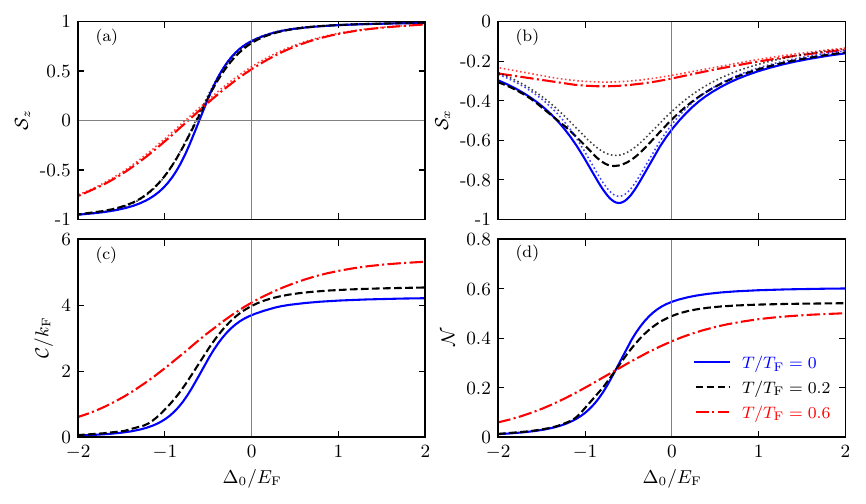}
    \caption{ \replyadd{The magnetization $\mathcal{S}_z$ (a), in-plane magnetization $\mathcal{S}_x$ (b), contact (c), and number of atoms in the dressing cloud (d),} are plotted as a function of detuning $\Delta_0/\ef $ for interaction $1/\kf\as=0$, Rabi coupling $\Omega_0=0.5\ef$, and temperatures $T/T_{\rm F}=0$ (blue solid), $T/T_{\rm F}=0.2$ (black dashed), and $T/T_{\rm F}=0.6$ (red dot-dashed). The dotted lines for the magnetizations correspond to Eq.~\eqref{eq:Sweak} where for each temperature we have calculated the polaron energy and quasiparticle residue in the absence of Rabi coupling \cite{Note1}.
    }
    \label{fig:comp_therm}
\end{figure*}

\begin{figure}
    \centering
    \includegraphics[width=0.45\textwidth]{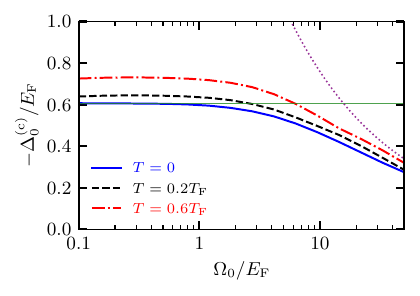}
    \caption{The critical detuning $\Delta_0^{\rm(c)}$ where $\mathcal{S}_z=0$ at unitarity $1/\kf\as=0$,
    as a function of the Rabi coupling $\Omega_0/\ef$, %
    for temperatures $T=0$ (blue solid), $T=0.2 T_{\rm F}$ (black dashed), and  $T=0.6 T_{\rm F}$ (red dot-dashed). The dotted purple line is the analytical result obtained %
    from the strong Rabi-coupled limit, Eq.~\eqref{eq:Delta-c-large}, 
    and the solid green line is the zero-temperature polaron energy $\mathcal{E}_0 \simeq -0.61 \ef$
    in the absence of a Rabi drive. 
    }
    \label{fig:magzero_uni}
\end{figure}

The changing spin composition of the ground state is reflected in the thermodynamic variables associated with interactions, namely the contact and the number of particles in the dressing cloud. We find that both of these increase from 0 to their values at unitarity in the absence of Rabi coupling: $\mathcal{C}_0/k_F=4.28$ \cite{Punk2009} and $\mathcal{N}_0=-{\mathcal E}_0/\ef=0.61$ \cite{Massignan2012}. For the number of particles in the dressing cloud, we can understand this by considering the expression for the free energy at unitarity and $T=0$ in Eq.~\eqref{eq:freeenergyunitarityt0}, which links the number of particles in the dressing cloud to $\mathcal{S}_z$ and $\mathcal{S}_x$. By using the expressions for the magnetization parameters at weak Rabi drive in Eq.~\eqref{eq:Sweak0}, we find
\begin{align} \label{eq:Nweak0}
    {\mathcal N}&= \frac12\left(1+\frac{\delta}{\sqrt{\Omega^2+\delta^2}}\right){\mathcal N}_0.
\end{align}
This is seen to closely match our numerical results in Fig.~\ref{fig:comp_T0}(d). Intuitively, this expression can be understood as being the product of the spin $\up$ fraction in the ground state (i.e., the interacting part of the impurity) and the number of particles in the dressing cloud in the absence of Rabi coupling. In particular, at $\delta=0$ where the impurity is half $\up$ and half $\down$ we expect the number of particles in the dressing cloud to be halved due to the Rabi coupling.

The contact is less constrained than the number of particles in the dressing cloud, since it is not present in the free energy at unitarity---see Eq.~\eqref{eq:freeenergyunitarityt0}. \replyadd{%
However, it is then reasonable to assume that we have the following approximate form:}
\begin{align} \label{eq:Cweak0}
    {\mathcal C}&\simeq \frac12\left(1+\frac{\delta}{\sqrt{\Omega^2+\delta^2}}\right){\mathcal C}_0,
\end{align}
\replyadd{which corresponds to the $\up$ contact in the absence of Rabi coupling multipled by the spin-$\up$ fraction (since the non-interacting $\down$ state has zero contact).}
Indeed, this yields an excellent agreement at weak Rabi drive \replyadd{[see Fig.~\ref{fig:comp_T0}(c)]}, but it overestimates the contact when $\Omega_0\gtrsim\ef.$ \replyadd{We note that the contact can be calculated from Eqs.~\eqref{eq:free_weak} and \eqref{eq:cont_def}; however even at weak Rabi coupling we do not obtain a simple analytic form.}

Figure~\ref{fig:comp_therm} illustrates the effect of temperature on these results. A non-zero temperature allows excited states to be populated which tends to reduce both the $\mathcal{S}_z$ and $\mathcal{S}_x$ magnetizations towards zero. This behavior is well captured by the analytic expressions in  Eq.~\eqref{eq:Sweak}. Furthermore, in the absence of Rabi coupling, it has been experimentally demonstrated that the contact parameter can be enhanced with increasing temperature~\cite{Yan2019u},
a behavior which has been reproduced theoretically~\cite{Liu2020PRL}. We see a similar effect in the Rabi-coupled case for all detunings considered. On the other hand, the number of particles in the dressing cloud decreases with temperature when $\Delta_0\gtrsim -0.61\ef$, while increasing for detunings below the attractive polaron, which is again linked to thermally populating excited states.

The success of the relatively simple analytic weak-Rabi-drive expressions in describing the interplay of Rabi coupling and strong interactions likely arises from the fact that the $\up$ impurity at unitarity features a ``dark continuum'' above its $T=0$ ground state~\cite{Goulko2016}, with strongly suppressed spectral weight. This implies that there are no nearby states that can strongly influence the impurity thermodynamics. In this sense, the experimental verification of our results would provide further strong evidence of the existence of a dark continuum. We also note that close to the polaron-molecule transition~\cite{Prokofev2008,Punk2009}, we expect significant spectral weight at energies comparable to the polaron, and hence we would expect stronger deviations from the weak-drive expressions.

Finally, Fig.~\ref{fig:magzero_uni} shows how the critical detuning $\Delta_0^{(\mathrm{c})}$ at which the magnetization $\mathcal{S}_z=0$ changes as a function of Rabi drive and temperature. As $\Omega_0\to0$ and $T\ll\ef$, the critical detuning is to a high degree of accuracy given by the polaron energy, calculated according to Eq.~\eqref{eq:epolup}. The polaron is known to initially shift to lower energies with increasing temperature, before eventually moving closer to zero energy~\cite{Hu2018,Tajima2019,Mulkerin2019}, as has been observed experimentally~\cite{Yan2019}. This is consistent with our findings, where we additionally observe the critical detuning to remain relatively constant until $\Omega_0\sim\ef$ before eventually moving towards 0. At very large Rabi drive, our results are consistent with the strong-drive results in Sec.~\ref{sec:strong}.

\section{Dynamics of the driven system}
\label{sec:dynamics}

We now turn to the dynamics of the Rabi-driven Fermi polaron. This is typically probed by measuring the magnetization as a function of time, as done previously in several Fermi-polaron experiments~\cite{Kohstall2012,Scazza2017,Oppong2019}. The dynamics of the Rabi-driven Fermi polaron has been theoretically described using a variational truncated basis method, both at $T=0$ \cite{Parish2016} and at finite temperature \cite{Adlong2020,Adlong2021}, and this has been successfully used to model experiments~\cite{Scazza2017,Oppong2019} focused on the repulsive branch~\cite{Adlong2020}. More recently, the problem has also been investigated using an approximate version of a many-body $T$-matrix approach~\cite{hu2022Rabi}, and a kinetic equation approach~\cite{wasak2022decoherence}.

\subsection{\replyadd{Dynamics in the time domain}}

To investigate the Rabi-driven dynamics, we imagine that the impurity is initially  in the non-interacting spin-$\down$ state at time $t=0$. Then we quantify the magnetization $\mathcal{S}_z$ at times $t\geq0$ via the spin-$\down$ fraction
\begin{align}
    N_{\down}(t) =\frac{\sum_{\p} e^{-\beta\epsilon_{\p}} N_{\p\down}(t)}{\sum_{\p} e^{-\beta\epsilon_{\p}}}. 
\end{align}
The Boltzmann average arises since we consider a single impurity (or, equivalently, a Boltzmann gas of uncorrelated impurities). The spin-$\down$ fraction for an impurity initially at momentum $\p$ is given by
\begin{align} \label{eq:Npdownt}
    N_{\p\down}(t) = \langle \hat{c}_{\p\downarrow}(t) \hat{n}_{\down} \hat{c}_{\p\downarrow}^{\dagger}(t)  \rangle,
\end{align}
where the expectation value is with respect to the states of the medium as in Eq.~\eqref{eq:Gt}. We have also used the time-dependent operator $\hat{c}^{\phantom{\dagger}}_{\p\down}(t) = e^{i\hat{H}t} \hat{c}^{\phantom{\dagger}}_{\p\down} e^{-i\hat{H}t}$ and we have defined  $\hat{n}_{\sigma} = \sum_{\p}\hat{c}^{\dagger}_{\p\sigma} \hat{c}_{\p\sigma}$.

We now introduce a new method to calculate the Rabi dynamics which involves expanding the expectation value in Eq.~\eqref{eq:Npdownt} in terms of excitations of the Fermi sea. This is equivalent to previous truncated basis calculations by some of the present authors~\cite{Parish2016,Adlong2020}, but the advantage of expanding the expectation value directly is that we can relate it to correlation functions, which in turn allows us to accurately describe the dynamics for much longer evolution times. We perform such an expansion by inserting a complete set of medium-only states between the impurity creation and annihilation operators in $\hat n_\down$ and restricting our attention to one excitation of the medium. This gives
\begin{alignat}{1}
&N_{\p\down}(t) = 
\expval*{ \hat{c}_{\p\downarrow}(t) 
\hat{c}_{\p\downarrow}^{\dagger}  }
\expval*{ \hat{c}_{\p\downarrow}
\hat{c}_{\p\downarrow}^{\dagger}(t)  } + \nonumber \\
\,& \sum_{\k\neq\q} \frac{\expval*{ \hat{c}_{\p\downarrow}(t) 
\hat{c}_{\p+\q-\k\downarrow}^{\dagger}\hat{f}_\k^{\dagger}\hat{f}_\q }\expval*{ \hat{f}_\q^{\dagger}\hat{f}_\k \hat{c}_{\p+\q-\k\downarrow} 
\hat{c}_{\p\downarrow}^{\dagger}(t) }}{\expval*{\hat{f}_\q^{\dagger}\hat{f}_\k \hat{f}_\k^{\dagger}\hat{f}_\q}} +\dots,
\label{eq:Ndowntexpand}
\end{alignat}
where we have applied Wick's theorem along with momentum conservation. The denominator in the last term ensures proper normalization of the inserted intermediate state.

The first-order term of Eq.~\eqref{eq:Ndowntexpand} is directly related to the spin-$\down$ time-dependent interacting Green's function: 
\begin{align}\label{eq:Ndown1}
    N^{(1)}_{\p\down}(t)=|G_\down(\p,t)|^2,
\end{align}
which can in principle be determined from the Fourier transform of Eq.~\eqref{eq:G_down}. However, in practice, we only require the spectral function $A_\down(\p,\omega)$, corresponding to the imaginary part of the Green's function, which we obtain by extending the time domain to $t<0$ and invoking time reversal symmetry. Thus, we can equivalently write Eq.~\eqref{eq:Ndown1} as~\cite{Adlong2020}
\begin{align}
    N^{(1)}_{\p\down}(t)=\int d\omega \, d\omega'A_\down(\p,\omega)A_\down(\p,\omega')e^{-i(\omega-\omega')t}.
\end{align}
This can be viewed as the ``bare'' impurity's contribution to the Rabi oscillations, as discussed in Ref.~\cite{Adlong2020}. However, at finite temperature, the impurity Green's function always decays to zero in the long-time limit~\cite{Adlong2020}; therefore this contribution alone is insufficient to describe the steady state of the Rabi-driven impurity.

\begin{figure}
    \centering
    \includegraphics[width=0.45\textwidth]{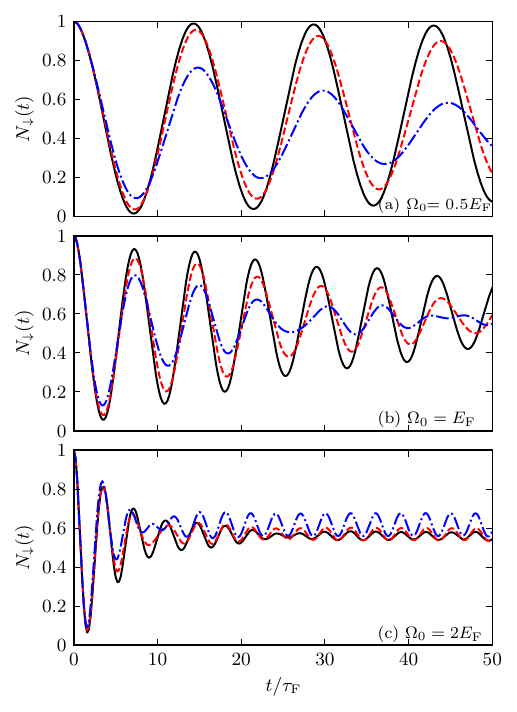}\centering
    \caption{Rabi oscillations at unitarity $1/\kf\as=0$ for different Rabi couplings (a) $\Omega_0=0.5\ef $, (b) $\Omega_0=\ef $, and (c) $\Omega_0=2\ef $, and at temperatures $T=0.03T_{\rm F}$ (black solid), $T=0.1T_{\rm F}$ (red dot-dashed), and $T=0.2T_{\rm F}$ (blue dashed). 
    For each temperature, we set the detuning equal to the attractive polaron energy in the absence of Rabi coupling \cite{Note1}.
    }
    \label{fig:rabidynamics}
\end{figure}

To capture the full dynamics, we need to include the contribution from the polaron dressing cloud, which is contained in the second-order term in Eq.~\eqref{eq:Ndowntexpand}: 
\begin{align} \label{eq:2ndorddyn}
N^{(2)}_{\p\down}(t)=\sum_{\k\neq\q} &
\frac{|\chi_\down(\k,\q ; \p,t)|^2}{n_\q (1-n_\k)}  .
\end{align} 
Here we have used $\expval*{\hat{f}_\q^{\dagger}\hat{f}_\k \hat{f}_\k^{\dagger}\hat{f}_\q} = n_\q (1-n_\k)$ and we have defined the spin-resolved two-body correlator
\begin{align}\label{eq:chisigmacorr}
\chi_{\sigma} (\k,\q; \p,t) = -i\theta(t) \, \langle \hat{c}_{\p\sigma}(t) \, \hat{c}_{\p+\q-\k\sigma}^{\dagger}\hat{f}_\k^{\dagger}\hat{f}_\q  \rangle.
\end{align}
Its Fourier transform $\chi_{\sigma} (\k,\q; \p, \omega)$ can be related to the impurity Green's function as follows (see Appendix \ref{app:2bodycorr} for details of the derivation):
\begin{alignat}{1} \nn
\chi_{\uparrow} (\k,\q; \p,\omega) =  \ & G_{\uparrow} ^{(\Omega)}(\p+\q-\k,\omega+\epsilon_\q-\epsilon _\k) \\ \label{eq:chiupfromGamma}
& \times \Gamma(\q; \p, \omega) \left(1-n_\k\right),
\end{alignat}
where $G_{\uparrow} ^{(\Omega)}(\p,\omega)$ is the non-interacting Green's function given in Eq.~\eqref{eq:G0}, and we have the interaction term
\begin{alignat}{1} \label{eq:GammafromTandG}
\Gamma(\q; \p, \omega) &= g \sum_\k \chi_{\uparrow} (\k,\q; \p,\omega) \nonumber \\
& = \mathcal{T}(\p+\q,\omega+\epsilon_\q) G_{\uparrow} (\p,\omega)n_\q, 
\end{alignat}
with $\mathcal{T}(\p,\omega)$ the medium $T$ matrix defined in Eq.~\eqref{eq:invT-matrix}. The above expressions assume that there is at most one excitation of the medium, consistent with the expansion in Eq.~\eqref{eq:Ndowntexpand}. Finally, we obtain the $\down$ two-body correlator relevant for Eq.~\eqref{eq:2ndorddyn}  using the exact relation between the $\uparrow$ and $\downarrow$ correlators:
\begin{align} \label{eq:chidownfromup}
\chi_{\downarrow}(\k,\q ; \p, \omega+\Delta_0) = & \frac{(\Omega_0^2/4)\chi_{\uparrow}(\k,\q ; \p, \omega+\Delta_0)}{(\omega - \epsilon_{\p}^{\phantom{\p}} )(\omega -\epsilon_{\k}-\epsilon_{\p+\q-\k}+\epsilon_{\q}) }. 
\end{align}

\begin{figure}
    \centering
    \includegraphics[width=0.45\textwidth]{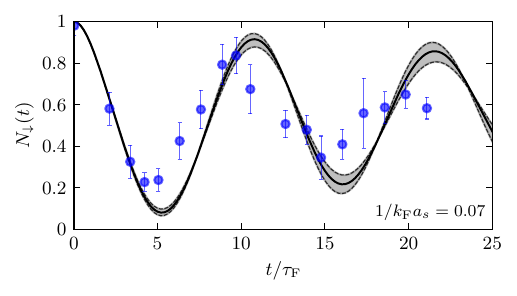}\centering
    \caption{Comparison between our numerical results     (solid black) and the experimental data for Rabi-driven $^6$Li near unitarity from Ref.~\cite{Scazza2017} (blue dots). The parameters are $\Omega_0=0.7\ef $,  $T\simeq0.14T_{\rm F}$ and     $1/\kf\as=0.07$. We set the detuning equal to the attractive polaron energy in the absence of Rabi drive, i.e., $\Delta_0\simeq     -0.683\ef$. The black dashed and shaded regions indicate the $20\%$ confidence interval for the experimental temperature. 
    }
    \label{fig:rabidynamics-expt}
\end{figure}

Figure \ref{fig:rabidynamics} displays the numerically calculated Rabi oscillations obtained within the one-excitation approximation. We fix the detuning to be resonant with the $\up$ attractive polaron such that $\mathcal{S}_z$ is close to zero in the steady state at unitarity (see Fig.~\ref{fig:magzero_uni}), corresponding to the long-time limit $N_\down(t\to \infty) \approx 0.5$.
We find that the initial time dependence during the first few oscillations is dominated by the first-order term $N^{(1)}_\down(t)$, with the gross features of the oscillations being dictated by the impurity spectral function, i.e., we find the Rabi frequency $\Omega = \sqrt{Z_0}\Omega_0$, in agreement with previous work~\cite{Kohstall2012,Adlong2020}. However, at longer times, the behavior becomes completely determined by the second-order ``dressing cloud'' term $N^{(2)}_\down(t)$ and, most notably, the spin-$\down$ fraction tends to something nonzero in the limit $t\to \infty$. The steady-state value is slightly shifted from the expected magnetization obtained from the free energy, indicating that higher order correlations are required to describe thermalization. However, our approximation is at least conserving since our numerical results satisfy the sum rule $N_\down(t) + N_\up(t) = 1$ (see Appendix \ref{app:sumrule}), which is not the case when only $N^{(1)}_\down(t)$  is considered~\cite{hu2022Rabi}. 

\begin{figure*}
    \centering
    \includegraphics[width=0.95\textwidth]{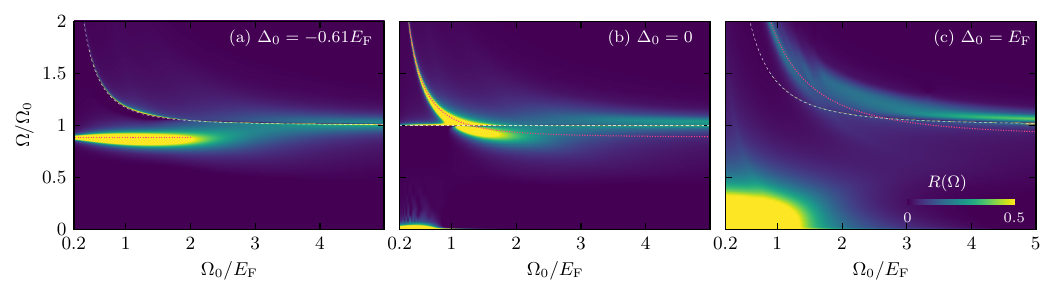}\centering
    \caption{\replyadd{Color plot of the} Rabi \replyadd{spectral function} $R(\Omega)$, corresponding to the Fourier transform of the Rabi oscillations, at interaction $1/\kf\as=0$, temperature $T=0.03T_{\rm F}$, and detunings (a) $\Delta_0= \mathcal{E}_0 =     -0.61E_{\rm F}$, (b) $\Delta_0=0$, and (c) $\Delta_0=E_{\rm F}$. The white dashed line is the Rabi frequency for a bare driven impurity, \replyadd{i.e.,} $\Omega = \sqrt{\Delta_0^2+\Omega_0^2}$, while the red dotted line shows the expected Rabi frequency for the attractive polaron, $\Omega = \sqrt{\left(\Delta_0-\mathcal{E}_0\right)^2+Z_0\Omega_0^2}$.
    }
    \label{fig:rabispectrum}
\end{figure*}

We also observe a damping of the oscillations that typically increases with temperature for the low temperatures in Fig.~\ref{fig:rabidynamics} and is larger than the finite-temperature damping rate $\Gamma_0$ of the attractive polaron~\cite{Note1}. Thus, the impurity momentum appears to play an important role in the damping of Rabi oscillations at finite temperature and its inclusion is necessary to capture the damping observed in experiments on $^6$Li~\cite{Scazza2017}, as shown in Fig.~\ref{fig:rabidynamics-expt}. This is in contrast to the case where the Rabi drive is resonant with the repulsive polaron, in which case the experiments could be successfully modelled with a zero-momentum variational approach~\cite{Adlong2020}. The difference between attractive and repulsive polarons is likely due to the processes that underlie their decoherence: in the former, momentum relaxation is expected to dominate~\cite{Bruun2008_coll}, while in the latter, there is ``many-body dephasing''~\cite{Adlong2020} which is present at zero momentum and temperature. \replyadd{Indeed, our results in Fig.~\ref{fig:rabidynamics-expt} qualitatively match those of Ref.~\cite{wasak2022decoherence}, which describes momentum relaxation within a quantum kinetic framework. Finally,} note that the experiment~\cite{Scazza2017}  \replyadd{is carried out in a harmonic trap, it features a finite density of impurities, and there are} 
weak interactions between the $\down$ impurities and the medium. \replyadd{All of these effects} are neglected in our theory, which could account for some of the discrepancies between theory and experiment in Fig.~\ref{fig:rabidynamics-expt}.

\subsection{\replyadd{Dynamics in the frequency domain}}

Further insight into the dynamics can be gained from the Fourier transform $R(\Omega) = \int dt N_\down(t) e^{i\Omega t}$, where we assume time-reversal symmetry of $N_\down(t)$. The result is depicted \replyadd{as a color map} in Fig.~\ref{fig:rabispectrum} and reveals the different frequencies present in the Rabi oscillations. At low Rabi drive $\Omega_0 \lesssim \ef$, the Rabi spectrum at low temperatures is dominated by the expected Rabi frequency for the attractive polaron, which is $\Omega = \sqrt{Z_0} \Omega_0$ at the resonance condition $\Delta_0 = \mathcal{E}_0$ [Fig.~\ref{fig:rabispectrum}(a)]. Moreover, we see that it continues to dominate even away from resonance in Fig.~\ref{fig:rabispectrum}(b,c), where the polaron Rabi frequency generalizes to $\Omega = \sqrt{\left(\Delta_0-\mathcal{E}_0\right)^2+Z_0\Omega_0^2}$ \replyadd{(red dotted line)}. However, once the detuning lies within the continuum of the repulsive branch, as in Fig.~\ref{fig:rabispectrum}(c), the oscillations are strongly damped and there is an overall decay of the spin-$\down$ fraction.  

With increasing Rabi drive in Fig.~\ref{fig:rabispectrum}, an additional sharp peak appears at the Rabi frequency for the bare driven impurity, i.e., $\Omega=\sqrt{\Delta_0^2+\Omega_0^2}$ \replyadd{(white dashed line)}. This arises from the non-interacting Green's function in Eq.~\eqref{eq:chiupfromGamma} and can be viewed as the oscillations of the high-momentum impurity states that make up the dressing cloud. This bare contribution is also apparent in Fig.~\ref{fig:rabidynamics}(c), where it manifests as small persistent oscillations at long times. In a real system, we expect such oscillations to be damped by higher order medium excitations, beyond what is included in our theory. However, it remains an interesting and open question whether one could observe a driven polaron evolving into a driven bare impurity with time, especially since our numerics suggest that the bare oscillations can be enhanced at higher temperatures.

\begin{figure}
    \centering
    \includegraphics[width=0.45\textwidth]{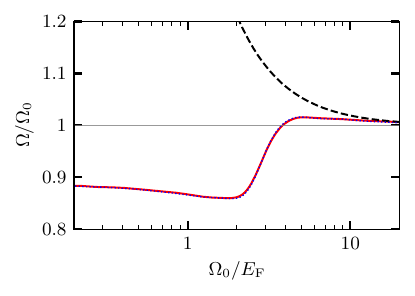}\centering
    \caption{Position of the peak in the Rabi spectrum at $1/\kf\as=0$ \replyadd{and temperature $T=0.03T_{\rm F}$}, for a detuning fixed to the attractive polaron energy, i.e., $\Delta_0=-0.61E_{\rm F}$ (red solid), and for the critical detuning $\Delta_0^{(c)}$ at zero magnetization in Fig.~\ref{fig:magzero_uni} (blue dotted). The black dashed line is the strong-Rabi-drive limit $\Omega=\Omega_0+\frac{4\pi n}{m^{3/2}\sqrt{\Omega_0}}$.
    }
    \label{fig:rabispectrum-peak}
\end{figure}

At strong Rabi drive $\Omega_0 \gtrsim \ef$, the Rabi frequency for the attractive polaron becomes strongly modified and approaches the Rabi drive frequency $\Omega_0$ as $\Omega_0 \to \infty$. We can better visualize this behavior by plotting the position of the main peak in Fig.~\ref{fig:rabispectrum}(a), as shown in Fig.~\ref{fig:rabispectrum-peak}. We see that the peak evolves in a rather non-monotonic manner with increasing $\Omega_0$ and it appears to jump from frequencies around $\sqrt{Z_0}\Omega_0$ to frequencies above $\Omega_0$. We can understand the latter behavior by considering the impurity quasiparticles that exist in the large-Rabi-drive limit. Specifically, if we identify the energies of the two Rabi-split quasiparticles, then the oscillation frequency is given by their energy difference.  
We have already obtained the energy of the lowest quasiparticle in Eq.~\eqref{eq:E-strong}; the energy of the excited quasiparticle is even simpler since the self energy is purely imaginary at lowest order and hence the interaction shift vanishes in the limit $\Omega_0 \to \infty$, thus reducing the quasiparticle energy to the single-particle energy: $E_\0^+ \simeq \Omega_0/2 + \Delta_0/2$. Therefore, the Rabi frequency is $\Omega=\Omega_0+\frac{4\pi n}{m^{3/2}\sqrt{\Omega_0}}$, which agrees with the strong-Rabi-drive limit in Fig.~\ref{fig:rabispectrum-peak}. This result is insensitive to the precise detuning $\Delta_0$ and indeed we see that this extends across the full range of $\Omega_0$ since the detunings $\Delta_0=\mathcal{E}_0$ and $\Delta_0^{(c)}$ give essentially indistinguishable results for the Rabi frequency. 

\section{Concluding remarks}
\label{sec:conc}

To conclude, we have investigated the thermodynamic and dynamic properties of a mobile spin-1/2 impurity immersed in a Fermi gas. For the thermodynamic properties that characterize both the interactions and the magnetization,
we have provided exact results in the limits of weak and strong Rabi drive compared with the Fermi energy. For a weak drive, these expressions depend on the properties of the addressed quasiparticle, which can be accurately determined either from experiment or using a wealth of approximation methods~\cite{Massignan2014,Scazza2022}. Our approach is based on calculating the impurity free energy~\cite{Liu2020PRL,Liu2020PRA}, which we obtain in the regime of an intermediate Rabi drive through the means of a many-body $T$-matrix approach. We find that our weak-Rabi-drive expressions are remarkably robust and extend up to intermediate Rabi drive, which might be due to the existence of a ``dark continuum'' in the impurity spectrum~\cite{Goulko2016}.

To extend our results to  dynamics, we have developed a Green's function method, which is formally equivalent to previous variational methods~\cite{Parish2016,Adlong2020}, but which has the advantage that it is numerically less involved, allowing us to accurately model the dynamics to much longer times. Here we have found that the inclusion of finite temperature and finite impurity momentum can lead to a significant damping of the attractive polaron Rabi oscillations, in agreement with experiment~\cite{Kohstall2012,Scazza2017}.

In the future, it will be interesting to investigate how a strong Rabi drive could affect the ground-state transition from a polaron quasiparticle to a state where the impurity binds a fermion to form a dressed molecule~\cite{Prokofev2008,Punk2009}. Since Rabi oscillations go beyond linear response and probe both the spectral function and higher-order correlations in the dressing cloud, they appear to be ideally suited to investigate the interplay between this single-particle transition and its associated many-body phase transition. In the absence of a Rabi drive, the polaron-molecule transition has been predicted to exist for the Fermi polaron at $1/\kf\as\simeq0.9$~\cite{Prokofev2008,Punk2009}. Rabi drives with $\Omega_0\gtrsim \ef,1/m\as^2$ thus yield the prospect of strongly modifying the diatomic molecule and hence of tuning the elusive quasiparticle transition. 

\acknowledgments 
We thank Nir Navon, Franklin Vivanco, and Alexander Schuckert for inspiring discussions, and Francesco Scazza for providing us with the experimental data from Ref.~\cite{Scazza2017}. BCM, JL, and MMP acknowledge support from the Australian Research Council
Centre of Excellence in Future Low-Energy Electronics Technologies
(CE170100039).  JL and MMP are also supported through the Australian Research
Council Future Fellowships FT160100244 and FT200100619, respectively, and JL through DP210101652.

\appendix

\section{Free energy}
\label{app:freeF}
In this Appendix we present the details of how to obtain the free energy relation, Eq.~\eqref{eq:freeenergy_rel}. In the single-impurity limit, the spectral function $A_\sigma(\p,\omega)$ coincides with the unoccupied spectral function~\cite{Haussmann2009}
\begin{align}\label{eq:Aplus}
    A_{\sigma+}\left(\p,\omega\right) &= \sum_{n,\nu} \frac{e^{-\beta E_{n}}}{Z_{\rm med}} |\langle n | \hat{c}_{\p\sigma}| \nu \rangle |^2 
    \delta\left( \omega + E_n -\mathcal{E}_{\nu} \right).
\end{align}
Here, we define the eigenenergies and states of the system (medium plus impurity) as $\mathcal{E}_{\nu}$ and $\ket{\nu}$, respectively, as well as those of the medium in the absence of the impurity, $E_{n}$ and $\ket{n}$. Then we have the medium partition function $Z_{\rm med} = \sum_{n}e^{-\beta E_{n}}$.

Analogously to Eq.~\eqref{eq:Aplus}, we have the occupied spectral function~\cite{Haussmann2009}
\begin{align} \label{eq:occupied}
A_{\sigma-}\left(\p,\omega\right) &= \sum_{n,\nu} \frac{e^{-\beta \mathcal{E}_{\nu}}}{Z_{\rm int}} |\langle n | \hat{c}_{\p\sigma}| \nu \rangle |^2 \delta\left(\omega + E_n -\mathcal{E}_{\nu} \right),
\end{align}
with corresponding partition function $Z_{\rm int} = \sum_{\nu}e^{-\beta \mathcal{E}_{\nu}}$. By using the properties of the delta functions, we find that the occupied and unoccupied spectral functions satisfy the detailed balance condition~\cite{Liu2020PRA}
\begin{align} \label{eq:det_bal}
A_{\sigma-}(\p,\omega)
& = e^{-\beta\omega} \frac{Z_{\rm med}}{Z_{\rm int}} A_{\sigma+}(\p,\omega).
\end{align}

To proceed, we make use of the sum rule,
\begin{alignat}{1} 
\int d\omega \, A_{\sigma-}\left(\p,\omega\right) &= n_{\rm int}^{\sigma}\left(\p\right),
\end{alignat}
which follows from the definition in Eq.~\eqref{eq:occupied}.
Here, $n_{\rm int}^{\sigma}\left(\p\right) = {\rm Tr}[ \hat{\rho} %
\hat{c}^{\dagger}_{\p\sigma}\hat{c}^{\phantom{\dagger}}_{\p\sigma} ]$ are the impurity momentum distribution densities of the interacting many-body system, and $\hat\rho$ the corresponding density matrix (see the discussion below Eq.~\eqref{eq:Sz}). Since we have a single impurity, we therefore arrive at
\begin{align} \label{eq:sum_rule}
    \int d\omega \sum_{\p,\sigma}\, A_{\sigma-}\left(\p,\omega\right)=1.
\end{align}

As discussed in Sec.~\ref{sec:thermodyn}, we define the impurity free energy as the difference between the free energy in the presence of both interactions and Rabi coupling, and the free energy of the system where the impurity is completely decoupled from the medium and there is no Rabi coupling. From the relationship between the free energy and the partition function, we therefore have the simple relation
\begin{align}
    e^{-\beta \mathcal{F}}=\frac{Z_{\mathrm{int}}}{Z_{\mathrm{med}}Z_{\mathrm{imp}}},
\end{align}
where $Z_{\mathrm{imp}}=\sum_\p e^{-\beta \epsilon_\p}$ is the partition function of a single impurity in the absence of Rabi coupling. Using the detailed balance condition \eqref{eq:det_bal} and the sum rule \eqref{eq:sum_rule}, we find
\begin{align}
    e^{-\beta \mathcal{F}}&=\frac{Z_{\rm int}}{Z_{\rm med}}\frac{\int d\omega \sum_{\p,\sigma}\, A_{\sigma-}\left(\p,\omega\right)}{\sum_\p e^{-\beta\epsilon_\p}} \nn \\
    & = \frac{\int d\omega \sum_{\p,\sigma}\, e^{-\beta\omega}A_{\sigma+}\left(\p,\omega\right)}{\sum_\p e^{-\beta\epsilon_\p}}.
\end{align}
The equivalence between $A_+$ and $A$ then yields Eq.~\eqref{eq:freeenergy_rel} in the main text.

We can use the same ideas to find the magnetization directly from the spectral function. We have
\begin{alignat}{1}
\mathcal{S}_z &= \frac{N_\up-N_\down}{N_\up+N_\down} \nn \\
&=\frac{\int d\omega \sum_{\p} \left[A_{\up-}\left(\p,\omega\right)-A_{\down-}\left(\p,\omega\right)\right]}{\int d\omega \sum_{\p} \left[A_{\up-}\left(\p,\omega\right)+A_{\down-}\left(\p,\omega\right)\right] } \nn \\
&=\frac{\int d\omega \,e^{-\beta\omega}\sum_{\p} \left[A_{\up}\left(\p,\omega\right)-A_{\down}\left(\p,\omega\right)\right]}{\int d\omega\, e^{-\beta\omega}\sum_{\p} \left[A_{\up}\left(\p,\omega\right)+A_{\down}\left(\p,\omega\right)\right] }.
\end{alignat}
However, we find that this formulation is not as numerically stable as first calculating the free energy and then taking the appropriate derivatives.

Finally, we note that we can straightforwardly extend the detailed balance relations to a finite density of impurities. This directly follows the arguments of Ref.~\cite{Liu2020PRA} (see Appendix D of that work) which we do not repeat here.

\begin{widetext}
\section{Two-body correlator} \label{app:2bodycorr}

In this appendix, we derive the spin-resolved two-body correlators used to calculate the full dynamics of the dressing cloud in the Rabi oscillations. We start by considering the Fourier transform of the spin-$\sigma$ impurity Green's function
\begin{align}
    G_\sigma (\p, \omega) = \expval{\hat{c}_{\p\sigma} \frac{1}{\omega -\hat{H}+i0} \hat{c}^\dag_{\p\sigma}} \, ,
\end{align}
where the average is over medium-only states and the Hamiltonian measures energy from that of the medium state in the absence of the impurity. 
Using the fact that we can rewrite the inverse operator as
\begin{align} \label{eq:expansion}
    \frac{1}{\omega -\hat{H}+i0} = \frac{1}{\omega -\hat{H}_0-\hat{H}_\Omega+i0} + \frac{1}{\omega -\hat{H}+i0} \hat{V} \frac{1}{\omega -\hat{H}_0-\hat{H}_\Omega+i0} ,
\end{align}
we have
\begin{align}
    G_\up ( \p, \omega) & =  \expval{\hat{c}_{\p\up} \frac{1}{\omega -\hat{H}_0-\hat{H}_\Omega+i0} \hat{c}^\dag_{\p\up}} + \expval{\hat{c}_{\p\up} \frac{1}{\omega -\hat{H}+i0} \hat{V} \frac{1}{\omega -\hat{H}_0-\hat{H}_\Omega+i0} \hat{c}^\dag_{\p\up}} \\
    & =  G_\up^{(\Omega)} (\p,\omega) + g \sum_{\k,\k',\q} \expval{\hat{c}_{\p\up} \frac{1}{\omega -\hat{H}+i0} \hat{c}^\dag_{\k\up}  \hat{f}^\dag_{\k'} \hat{f}_{\k'-\q} \hat{c}_{\k+\q\up} \frac{1}{\omega -\hat{H}_0-\hat{H}_\Omega+i0} \hat{c}^\dag_{\p\up}} \\
    & = G_\up^{(\Omega)} (\p,\omega) + g \sum_{\k,\q} \expval{\hat{c}_{\p\up} \frac{1}{\omega -\hat{H}+i0} \hat{c}^\dag_{\p+\q-\k\up}  \hat{f}^\dag_{\k} \hat{f}_{\q}}
    \expval{\hat{c}_{\p\up} \frac{1}{\omega -\hat{H}_0-\hat{H}_\Omega+i0}  \hat{c}^\dag_{\p\up}} .
\end{align}
In the last line, we have used Wick's theorem as well as the fact that the single-particle Hamiltonian $\hat{H}_0+\hat{H}_\Omega$ does not change the impurity momentum. Defining the spin-$\up$ two-body correlator
\begin{align} \label{eq:chiup_FT}
    \chi_\up (\k,\q;\p,\omega) = \expval{\hat{c}_{\p\up} \frac{1}{\omega -\hat{H}+i0} \hat{c}^\dag_{\p+\q-\k\up}  \hat{f}^\dag_{\k} \hat{f}_{\q}}_{\k \neq \q},
\end{align}
the spin-$\uparrow$ impurity Green's function can be written as
\begin{align}
    \label{eq:Gupexpand}
    G_\up (\p,\omega) & = G_\up^{(\Omega)} (\p,\omega) \left[ 1 + g \sum_\q \! \expval*{\hat{f}_{\q}^\dag \hat{f}_{\q}} G_\up (\p,\omega) + g \sum_{\k,\q} \chi_\up (\k,\q;\p,\omega) \right] \, .
\end{align}
The two-body correlator, Eq.~\eqref{eq:chiup_FT}, is the Fourier transform of Eq.~\eqref{eq:chisigmacorr}.

The spin-$\up$ two-body correlator can now be expanded by inserting Eq.~\eqref{eq:expansion}
\begin{align}
    \chi_\up (\k,\q;\p,\omega) = \expval{\hat{c}_{\p\up} \frac{1}{\omega -\hat{H}+i0} \hat{V} \frac{1}{\omega -\hat{H}_0-\hat{H}_\Omega+i0} \hat{c}^\dag_{\p+\q-\k\up}  \hat{f}^\dag_{\k} \hat{f}_{\q}}_{\k \neq \q} \, ,
\end{align}
where the first-order term is zero since we require $\k \neq \q$. Performing the different contractions for the $\hat{f}$ operators, we then obtain
\begin{align} \nn
    \chi_\up (\k,\q;\p,\omega) & = g \sum_{\q'} \! \expval*{\hat{f}_{\q'}^\dag \hat{f}_{\q'}} \expval{\hat{c}_{\p+\q-\k\up} \frac{1}{\omega - E^\p_{\k\q}- \hat{H}_\Omega +i0} \hat{c}^\dag_{\p+\q-\k\up}} \chi_\up (\k,\q;\p,\omega)  \\ \nn
    & + g \expval*{\hat{f}_{\q}^\dag \hat{f}_{\q}} \expval*{\hat{f}_{\k} \hat{f}^\dag_{\k}}
    \expval{\hat{c}_{\p+\q-\k\up} \frac{1}{\omega - E^\p_{\k\q}- \hat{H}_\Omega +i0} \hat{c}^\dag_{\p+\q-\k\up}} G_\up (\p,\omega) \\ \nn
    & + g  \expval*{\hat{f}_{\k} \hat{f}^\dag_{\k}} \expval{\hat{c}_{\p+\q-\k\up} \frac{1}{\omega - E^\p_{\k\q}- \hat{H}_\Omega +i0} \hat{c}^\dag_{\p+\q-\k\up}} \sum_{\k'} \chi_\up (\k',\q;\p,\omega) \\
    & - g \expval*{\hat{f}_{\q}^\dag \hat{f}_{\q}} \expval{\hat{c}_{\p+\q-\k\up} \frac{1}{\omega - E^\p_{\k\q}- \hat{H}_\Omega +i0} \hat{c}^\dag_{\p+\q-\k\up}} 
    \sum_{\q'}  \chi_\up (\k,\q';\p,\omega) \, ,
    \label{eq:chiupexpand}
\end{align} 
where $E^\p_{\k\q} = \epsilon_{\p+\q-\k} + \ek  -\eq$. 
Using the fact that
\begin{align}
    G^{(\Omega)}_\up (\p+\q-\k,\omega + \eq - \ek) = \expval{\hat{c}_{\p+\q-\k\up} \frac{1}{\omega - E^\p_{\k\q}- \hat{H}_\Omega +i0} \hat{c}^\dag_{\p+\q-\k\up}} \, ,
\end{align}
we finally obtain the coupled equations
\begin{subequations} \label{eq:ChevyG}
\begin{align} \label{eq:ChevyG-a}
     \chi_\up (\k,\q;\p,\omega) & = G^{(\Omega)}_\up ( \p+\q-\k,\omega + \eq - \ek) 
     \left[%
     g \expval*{\hat{f}_{\q}^\dag \hat{f}_{\q}} \expval*{\hat{f}_{\k} \hat{f}^\dag_{\k}} G_\up (\p, \omega) 
    + g  \expval*{\hat{f}_{\k} \hat{f}^\dag_{\k}}  \sum_{\k'} \chi_\up(\k',\q;\p, \omega) \right] \\
    \label{eq:ChevyG-b}
 G_\up (\p, \omega) & = G_\up^{(\Omega)} (\p, \omega) \left[ 1 + g \sum_{\k,\q} \chi_\up (\k,\q;\p,\omega) \right] \, .
\end{align}
\end{subequations}
In arriving at Eq.~\eqref{eq:ChevyG-a}, we have dropped the first and last lines of Eq.~\eqref{eq:chiupexpand} since both vanish when we take $g \to 0$ to renormalize the interactions. Similarly, in Eq.~\eqref{eq:ChevyG-b} we have dropped the Hartree term from Eq.~\eqref{eq:Gupexpand} which also renormalizes to zero. Note that the first term in Eq.~\eqref{eq:ChevyG-a} will also eventually renormalize to zero; however since it is the only term linking $G_\up$ to $\chi_\up$ we must keep it while we are considering the coupling between these quantities~\cite{Chevy2006}.

It is straightforward to show that this gives the expected impurity Green's function within the non-self-consistent $T$-matrix approximation. Specifically, if we define 
\begin{align}
    \Gamma(\q;\p, \omega) \equiv g \sum_{\k} \chi_\up (\k,\q; \p, \omega) ,
\end{align}
then performing a sum over $\k$ in Eq.~\eqref{eq:ChevyG-a} gives
\begin{equation}
    \left[\frac{1}{g} - \sum_\k \expval*{\hat{f}_{\k} \hat{f}^\dag_{\k}} G^{(\Omega)}_\up (\p+\q-\k,\omega + \eq - \ek) \right] \Gamma(\q;\p, \omega) = \expval*{\hat{f}_{\q}^\dag \hat{f}_{\q}} G_\up (\p, \omega) ,
\end{equation}
where we have used the fact that $g \sum_\k \expval*{\hat{f}_{\k} \hat{f}^\dag_{\k}} G^{(\Omega)}_\up (\p+\q-\k,\omega + \eq - \ek) \to 1$ as $\Lambda \to \infty$. The quantity in brackets is precisely the inverse $T$ matrix in the presence of Rabi drive, and thus we have Eq.~\eqref{eq:GammafromTandG} in the main text:
\begin{align}
   \Gamma(\q; \p, \omega)
& = \mathcal{T}(\p+\q,\omega+\epsilon_\q) G_{\uparrow} (\p,\omega)n_\q.
\end{align}
Rearranging for $\Gamma(\q;\p, \omega)$ and inserting this into Eq.~\eqref{eq:ChevyG-b} finally gives
\begin{align}
    G_\up (\p,\omega) & = G_\up^{(\Omega)} (\p,\omega) + G_\up^{(\Omega)} (\p,\omega) \underbrace{\sum_{\q} %
    n_\q\mathcal{T}(\p+\q,\omega+\epsilon_\q) }_{\Sigma(\p,\omega)}
    G_\up (\p, \omega)  .
\end{align}
This is precisely Dyson's equation for the impurity Green's function, with the self energy in Eq.~\eqref{eq:self-energy}.
We are now free to take the limit $g\to0$ in Eq.~\eqref{eq:ChevyG-a}, and thus we have Eq.~\eqref{eq:chiupfromGamma} from the main text:
\begin{alignat}{1}
\chi_{\uparrow} (\k,\q;\p,\omega) = G_{\uparrow} ^{(\Omega)}(\p+\q-\k,\omega+\epsilon_\q-\epsilon _\k) \Gamma(\q;\p,\omega) \left(1-n_\k\right).
\end{alignat}

To obtain the spin-$\down$ two-body correlator,
\begin{align} \label{eq:G2down}
    \chi_\down (\k,\q;\p,\omega) = \expval{\hat{c}_{\p\down} \frac{1}{\omega -\hat{H}+i0} \hat{c}^\dag_{\p+\q-\k\down}  \hat{f}^\dag_{\k} \hat{f}_{\q}}_{\k \neq \q}, 
\end{align}
we use the exact expansion
\begin{align} \notag
    \frac{1}{\omega -\hat{H}+i0} = & \frac{1}{\omega -\hat{H}_0-\hat{V}+i0} + 
    \frac{1}{\omega -\hat{H}_0-\hat{V}+i0} \hat{H}_\Omega \frac{1}{\omega -\hat{H}_0-\hat{V}+i0} \\ 
    & +  \frac{1}{\omega -\hat{H}_0-\hat{V}+i0} \hat{H}_\Omega \frac{1}{\omega -\hat{H}+i0} \hat{H}_\Omega \frac{1}{\omega -\hat{H}_0-\hat{V}+i0} ,
\end{align}
to relate it to $\chi_\up (\k,\q;\p,\omega)$. Inserting this expansion into Eq.~\eqref{eq:G2down} and using the fact that the first two terms vanish, we find the exact relation:
\begin{align}
    \chi_\down (\k,\q;\p,\omega) = \frac{\Omega_0^2/4}{(\omega - \ep -\Delta_0+i0) (\omega -E^\p_{\k\q} -\Delta_0+i0)} \chi_\up (\k,\q;\p,\omega) .
\end{align}
This is Eq.~\eqref{eq:chidownfromup} from the main text.

\end{widetext}

\section{Conservation of impurity number in Rabi dynamics}
\label{app:sumrule}

An important consistency check on our numerical calculation is that the total impurity number remains  conserved throughout the time evolution. That is, for an impurity initially at momentum $\p$ and time $t=0$, we require the following condition to be satisfied
\begin{align} \label{eq:Nupdownt}
    1 &= \expval*{\hat{c}_{\p\downarrow}(t) \left( \hat{n}_{\down} +\hat{n}_{\up} \right)  \hat{c}_{\p\downarrow}^{\dagger}(t)} \nonumber \\
    &= N_{\p\downarrow}(t) + N_{\p\up}(t),
\end{align} 
where we recall that %
$\hat{n}_{\sigma} = \sum_{\p}\hat{c}^{\dagger}_{\p\sigma} \hat{c}_{\p\sigma}$. The spin-$\downarrow$ fraction, $N_{\p\downarrow}(t) $, for an impurity
initially at momentum $\p$ is given by Eq.~\eqref{eq:Ndowntexpand}, and  by complete analogy we determine the spin-$\up$ fraction, $ N_{\p\up}(t)$,  at momentum $\p$ via 
\begin{alignat}{1}
&N_{\p\up}(t) = 
 \expval*{ \hat{c}_{\p\downarrow}(t) %
\hat{c}_{\p\uparrow}^{\dagger}  }
\expval*{ \hat{c}_{\p\uparrow}%
\hat{c}_{\p\downarrow}^{\dagger}(t)  }
+ \nonumber \\
\,& \sum_{\k\neq\q} \frac{\expval*{ \hat{c}_{\p\downarrow}(t) %
\hat{c}_{\p+\q-\k\downarrow}^{\dagger}\hat{f}_\k^{\dagger}\hat{f}_\q }\expval*{ \hat{f}_\q^{\dagger}\hat{f}_\k \hat{c}_{\p+\q-\k\uparrow} %
\hat{c}_{\p\uparrow}^{\dagger}(t) }}{\expval*{\hat{f}_\q^{\dagger}\hat{f}_\k \hat{f}_\k^{\dagger}\hat{f}_\q}} +\dots,
\label{eq:Ntottexpand}
\end{alignat}
where we have used Wick's theorem and momentum conservation.

The first order term in Eq.~\eqref{eq:Ntottexpand} is directly related to the off-diagonal Green's function in Eq.~\eqref{eq:green}:
\begin{alignat}{1} \label{eq:dens_time_ud}
N_{\p\up}^{(1)}(t) = |G_{\up\down}(\p,t)|^2,
\end{alignat}
where
\begin{align}
G_{\up\down}(\p,\omega) = \frac{\Omega_0/2}{\Omega_0^2/4 -\left(\omega-\epsilon_\p-\Delta_0 \right) \left( \omega-\epsilon_\p-\Sigma(\p,\omega) \right) }.
\end{align}
As in the main text, we utilize the time reversal symmetry of the Green's function to equivalently write Eq.~\eqref{eq:dens_time_ud} in terms of the spectral function $A_{\up\down}(\p,\omega)=-\frac{1}{\pi}{\rm Im}G_{\up\down}(\p,\omega)$:
\begin{align}
    N^{(1)}_{\p\up}(t)=\int d\omega \, d\omega'A_{\up\down}(\p,\omega)A_{\up\down}(\p,\omega')e^{-i(\omega-\omega')t}.
\end{align}
We note that the bare contributions, $N^{(1)}_{\p\down}(t)+N^{(1)}_{\p\up}(t)$, to Eq.~\eqref{eq:Nupdownt} do not conserve impurity number %
at finite times for any temperature.

To satisfy the single-impurity condition, we need to include the contribution from the polaron dressing cloud, which is contained in the second-order term in Eq.~\eqref{eq:Ntottexpand}: 
\begin{align} \label{eq:2ndorddyn_ud}
N^{(2)}_{\p\up}(t)=\sum_{\k\neq\q} &
\frac{|\chi_{\up\down}(\k,\q ; \p,t)|^2}{n_\q (1-n_\k)}  ,
\end{align}
where we define the  spin-$\up\down$ two-body  correlator as
\begin{align} %
    \chi_{\up\down} (\k,\q;\p,t) =  -i\theta(t)\expval*{\hat{c}_{\p\down}(t) \hat{c}^\dag_{\p+\q-\k\up}  \hat{f}^\dag_{\k} \hat{f}_{\q}}. 
\end{align}

The Fourier transform of the spin-$\up\down$ two-body  correlator can be related to the $\uparrow$  correlator $\chi_\up (\k,\q;\p,\omega)$
by using the expansion
\begin{align} \notag
    \frac{1}{\omega -\hat{H}+i0} = & \frac{1}{\omega -\hat{H}_0-\hat{V}+i0} + \nonumber \\
    &\frac{1}{\omega -\hat{H}_0-\hat{V}+i0} \hat{H}_\Omega \frac{1}{\omega -\hat{H}+i0}.
\end{align}
The first term does not contribute to the two-body correlator, and thus we have
\begin{align}
    \chi_{\up\down} (\k,\q;\p,\omega) = \frac{\Omega_0/2}{\omega - \ep -\Delta_0+i0 }\, \chi_\up(\k,\q;\p,\omega).
\end{align}
We numerically calculate Eq.~\eqref{eq:Nupdownt} as a function of time for all results presented in the paper. We find that it is always unity within a numerical error of $\lesssim 1\%$, and hence the sum rule is always satisfied in our numerics.

\bibliography{polaron}

\end{document}